\documentclass[showpacs,amsmath,amssymb,twocolumn,pra,superscriptaddress,notitlepage]{revtex4-1}

\usepackage[dvips]{graphicx}
\usepackage{subfigure}
\usepackage{amsfonts}
\usepackage{amssymb}
\usepackage{amscd}
\usepackage{amsmath}    
\usepackage{enumerate}
\usepackage{epsfig}
\usepackage{subfigure}
\usepackage{bm}
\usepackage{xcolor}
\usepackage{amsthm}
\usepackage{framed}
\usepackage{multirow}
\usepackage{mathrsfs,amssymb}
\usepackage{mathtools}
\usepackage{lipsum}

\newtheorem{theorem}{Theorem}

\newcommand{\bra}[1]{\mbox{$\left\langle #1 \right|$}}
\newcommand{\ket}[1]{\mbox{$\left| #1 \right\rangle$}}
\newcommand{\braket}[2]{\mbox{$\left\langle #1 \;\middle|\; #2 \right\rangle$}}
\newcommand{\tr}{\mathrm{tr}}

\begin{document}
\title{Numerical Method for Finite-size Security Analysis of Quantum Key Distribution}

\begin{abstract}
Quantum key distribution (QKD) establishes secure links between remote communication parties. As a key problem for various QKD protocols, security analysis gives the amount of secure keys regardless of the eavesdropper's computational power, which can be done both analytically and numerically. Compared to analytical methods which tend to require techniques specific to each QKD protocol, numerical ones are more general since they can be directly applied to many QKD protocols without additional techniques. However, current numerical methods are carried out based on some assumptions such as working in asymptotic limit and collective attacks from eavesdroppers. In this work, we remove these assumptions and develop a numerical finite-size security analysis against general attacks for general QKD protocols. We also give an example of applying the method to the recent Phase-Matching QKD protocol with a simple protocol design. Our result shows that the finite-size key rate can surpass the linear key-rate bound in a realistic communication time.
\end{abstract}

\author{Hongyi Zhou}
\affiliation{Photon Science Center, Graduate School of Engineering, The University of Tokyo, 7-3-1 Hongo, Bunkyo-ku, Tokyo 113-8656, Japan}
\affiliation{Institute of Computing Technology, Chinese Academy of Sciences, 100190, Beijing, China.}

\author{Toshihiko Sasaki}
\affiliation{Department of Applied Physics, Graduate School of Engineering, The University of Tokyo, 7-3-1 Hongo, Bunkyo-ku, Tokyo 113-8656, Japan}
\author{Masato Koashi}
\affiliation{Photon Science Center, Graduate School of Engineering, The University of Tokyo, 7-3-1 Hongo, Bunkyo-ku, Tokyo 113-8656, Japan}
\affiliation{Department of Applied Physics, Graduate School of Engineering, The University of Tokyo, 7-3-1 Hongo, Bunkyo-ku, Tokyo 113-8656, Japan}

\maketitle
\section{introduction}
Quantum key distribution (QKD) is one of the most successful applications of quantum mechanics and quantum information \cite{bennett1984quantum,gisin2002quantum,scarani2009security}. Guaranteed by the quantum no-cloning theorem, it enables the communication partners, Alice and Bob, to share keys with information-theoretic security, even if the eavesdropper has infinite computational power. Security analysis is the core issue of quantum key distribution. Currently, there exist two lines of security analysis methods, one is based on the phase error correction in an equivalent virtual QKD protocol, including Lo-Chau and Shor-Preskill's approach considering a virtual entanglement distillation protocol \cite{lo1999unconditional,shor2000simple}, and Koashi's approach considering the complementarity of virtual local measurement results \cite{koashi2009simple}; The other is an entropic framework \cite{Devetak2005Distillation,renner2008security}, which directly analyzes the correlation between the communication partners characterized by conditional min-entropy. In practical implementations, the effect of device imperfections \cite{gottesman2004security,Koashi2003BasisInd} and finite data size \cite{renner2008security,scarani2008quantum,tomamichel2012tight} on security are also proposed and well addressed.

All of the security analyses aim at finding a lower bound of the secret key rate, in other words, an upper bound of the information leakage to an eavesdropper. Conventional security analysis methods are analytical, where specific techniques are needed for different protocols. Recently, numerical methods for security analysis are proposed \cite{coles2016numerical, winick2018reliable, wang2019characterising,primaatmaja2019versatile}, based on convex optimization problems constrained by the statistics of measurement results in experiments. The target function of the convex optimization problems can be the phase error rate \cite{wang2019characterising,primaatmaja2019versatile} or conditional entropy \cite{coles2016numerical}, following the two lines of security analysis methods respectively. Both of them can be formulated \cite{wang2019characterising,primaatmaja2019versatile} or transformed \cite{winick2018reliable,lin2019asymptotic} to a semi-definite programming (SDP) problem. Recent researches show that the SDP method is successfully applied to  continuous variable QKD \cite{ghorai2019asymptotic,lin2019asymptotic} and device-independent QKD protocols \cite{tan2019computing}. For protocols with mixed state source, the SDP method is also applied to find an optimal twisting operation as well as the secure key rate \cite{bourassa2020loss}. 

Compared with analytical methods, numerical methods are more general, i.e., the same convex optimization problem framework can be applied to various protocols, even if the protocols are lack of symmetry \cite{coles2016numerical}. The asymmetry may come from the device imperfections in experiments, such as the source flaw and the misalignment error, or due to specific requirements to implement the protocols \cite{fung2006security}. In analytical methods, such asymmetry can be addressed by certain techniques, such as the quantum coin method dealing with the source flaw \cite{gottesman2004security,Lo2007Nonrandom}. In contrast, we usually do not need additional techniques in numerical methods. 
However, there are also restrictions for numerical methods. The convex optimization problem corresponds to a single round of quantum communication. As a result, most of the current numerical methods can only deal with the asymptotic security. 
Recently, there are two studies for the finite-size case that only carried out numerical analysis against collective attacks \cite{bunandar2020numerical,george2021numerical}. The work \cite{george2021numerical} also considers a possible extension to general attacks. Nevertheless, its performance is too pessimistic due to the direct application of the Finite Quantum de Finetti theorem \cite{renner2008security}.

In this work, we follow the phase error correction approach and develop a numerical finite-size analysis method against general attacks for a broad type of quantum key distribution protocols. Given the asymptotic analysis, we formulate a numerical framework for finite-size analysis by constructing virtual protocols. For the experiment outcomes, we introduce the nominal values and observed values, which are crucial in finite-size analysis. In contrast, such a distinction is not necessary in numerical asymptotic analyses. 
Considering the dual problem of the SDP given in \cite{wang2019characterising,primaatmaja2019versatile}, we can obtain an operator inequality for the phase error operator within one round of quantum communication, which is independent of the quantum state. By applying concentration inequalities \cite{azuma1967weighted,bernstein1924modification,kato2020concentration}, it is further transformed into an upper bound of the number of phase error occurrence in finite-size case under general attacks. One can directly substitute the experimental statistics into our framework and obtain secure finite-size key rates, without knowing any techniques of security analysis. To show how to apply our framework, we give an example of the recent Twin-Field QKD or Phase-Matching QKD \cite{lucamarini2018overcoming,Ma2018phase}, which are proved to surpass the linear key-rate bound \cite{takeoka2014fundamental,pirandola2017fundamental} in finite-size cases \cite{maeda2019repeaterless,lorenzo2019tight}. Our result shows such a result can still be achieved even with simplified protocol designs. 

\section{Numerical framework for prepare-and-measure QKD protocols}

\subsection{Numerical framework in asymptotic case}

We consider a broad type of prepare-and-measure protocols between the communication partners, Alice and Bob, which can be roughly described as the following process. Alice randomly prepares a quantum state and sends it to Bob. Bob also randomly performs a measurement on the incoming signal. After the measurement, they exchange some classical information about the state preparation and measurement result by an authenticated classical channel and probabilistically determine the round as a signal round or a test round. They evaluate the information leakage based on the results in the test rounds. With the classical information, they can perform the post-processing to extract secure keys.

The protocol can be described in the language of quantum information and mathematics. The normalized quantum state prepared by Alice is denoted as $\ket{\psi_i}\in \mathcal{H}_{A^\prime}$ with probability $\tau_A^i$ $(i\in\{0,1,2,\cdots,d_A\})$, where $\sum_i^{d_A} \tau_A^i =1$. Bob's measurement is given by a set of positive operator-valued measure (POVM) defined in $\mathcal{H}_{B}$, $\{\Gamma_y^{b}\}_b$, which is chosen with a probability $\tau_B^y$ and satisfies the normalization condition, $\sum_b \Gamma_y^{b} =I$ and $\sum_y \tau_B^y =1$. They repeat the process for $N_{\mathrm{tot}}$ rounds. Alice and Bob may randomly choose auxiliary random variables for each round, which are collectively denoted as $\alpha$ with probability $p_{\mathrm{aux},\alpha}$.
Alice and Bob publicly communicate a part of the results $(i,y,b,\alpha)$ abstractly denoted as $\tilde{\gamma}$, which is represented by a map $f_{\mathrm{ann}}$ from $(i,y,b,\alpha)$ to $\tilde{\gamma}$. They determine the role of each round based on $\tilde{\gamma}$ and a predetermined set $W$
such that $\tilde{\gamma}\in W$ for a signal round and $\tilde{\gamma} \notin W$ for a test round. 
In a signal round, Alice and Bob also generate their sifted-key bits based on $\tilde{\gamma}$ and information available to them.
In the case of direct reconciliation, Bob corrects his sifted key to match Alice's.
Hence we call Alice's sifted key at this point a reconciled sifted key.
The reconciled sifted-key bit $z\in\{0,1\}$ can be represented as a function $f_{\mathrm{rec}}(i,y,b,\alpha)$.
In the case of reverse reconciliation, Bob's sifted key becomes the reconciled sifted key.
In a test round, they make additional announcements to reveal the value of $(i,y,b,\alpha)$. 
We denote the probability that $\alpha$ belongs to a test round as
\begin{equation}
p_{\mathrm{test}}(i,y,b) = \sum_{\alpha:f_{\mathrm{ann}}(i,y,b,\alpha)\notin W }p_{\mathrm{aux},\alpha}.
\end{equation}
The dependence on $(i,y,b)$ is because a signal or test round is jointly determined by $(i,y,b,\alpha)$ in general. 
Based on the announcement in test rounds, they record the the number of test rounds with $(i,y,b)$ as $N^{i,y,b}_{\mathrm{test}}$ and the number of signal rounds $N_{\mathrm{sig}}$. 
Then they perform the post-processing to evaluate the information leakage to the eavesdropper and secure key rate.

Although the dimension of Bob's incoming signal can be infinite, we assume the dimension of $\mathcal{H}_{B}$ to be finite since we can only deal with finite-dimensional system in the numerical method.
A possible way of the finite-dimensional reduction is the squashing model \cite{Beaudry2008squashing}.
In general, such a model does not always exist for arbitrary QKD protocols.
We consider a more general case of tagging method \cite{gottesman2004security} in Appendix~\ref{app:tag}.

For analyzing the security of the protocol, it is convenient to clarify what is announced in each round. Since the value of $(i,y,b,\alpha)$ completely specifies what are announced by Alice and Bob, the distinct announcements are represented by disjoint sets of values of $(i,y,b,\alpha)$. More specifically, let us define the set of values of $(i,y,b,\alpha)$ leading to announcement $\tilde{\gamma} \in W$ by $\Omega_{\tilde{\gamma}}\coloneqq\{(i,y,b,\alpha): f_{\mathrm{ann}}(i,y,b,\alpha) =\tilde{\gamma} \in W\}$, which represents the announcement in a signal round. In a test round, the value of $(i,y,b,\alpha)$ is further announced in addition to $\tilde{\gamma} \notin W$, which corresponds to a single-element set $\{ (i,y,b,\alpha) \}$. All the possible announcements are thus represented by ${\cal W}_{\mathrm{all}}={\cal W} \cup {\cal W}_{\mathrm{test}}$, where
\begin{equation}
\begin{aligned}
{\cal W} &\coloneqq\{ \Omega_{\tilde{\gamma}} : \tilde{\gamma} \in W \} \\
{\cal W}_{\mathrm{test}} &\coloneqq\{ \{ (i,y,b,\alpha) \} : f_{\mathrm{ann}}( i,y,b,\alpha) \notin W \}.
\end{aligned}
\end{equation}
In this notation, the whole announcement in a round is represented by an element $\gamma \in {\cal W}_{\mathrm{all}}$, that is to say, Eve's receiving actual communication between Alice and Bob is information-theoretically equivalent to her receiving $\gamma$. Here $\gamma \in {\cal W}$ implies a signal round, and $\gamma \notin {\cal W}$ implies a test round. A value of $(i,y,b,\alpha)$ leads to announcement $\gamma$ if and only if $(i,y,b,\alpha) \in \gamma$.

An entanglement-based protocol is often considered for the convenience of security analysis, whose outputs (the classical-classical-quantum state shared by Alice, Bob and the eavesdropper Eve) are identical to those of the corresponding actual protocol. In the protocol, Alice prepares an entangled quantum state 
\begin{equation}
\label{eq:input}
\hat{\rho}_{\mathrm{in}} =\sum_{i,i^\prime} \sqrt{\tau_A^i \tau_A^{i^\prime}} \ket{i}\bra{i^\prime}_A \otimes \ket{\psi_i} \bra{\psi_{i^\prime}}_{A^\prime}.
\end{equation}
She keeps the $d_A$-dimensional ancillary system $\mathcal{H}_{A}$ and sends the signal part in $\mathcal{H}_{A^\prime}$ to Bob. 
They repeat the process for $N_{\mathrm{tot}}$ rounds, and share a bipartite state in $\mathcal{H}_{A}^{\otimes N_{\mathrm{tot}}} \otimes \mathcal{H}_{B}^{\otimes N_{\mathrm{tot}}}$. We denote the averaged bipartite state in one round as $\hat{\rho}_{\mathrm{out}}$. 
They perform the classical post-processing with the experimental results of $N_{\mathrm{tot}}$ rounds. 

In the security proofs based on the phase error correction, we define the phase error in the following way.
We first represent the process of each round in the actual protocol as a POVM in the entanglement-based protocol.
Its output in each round includes whether the round is signal or test, the bit $z$ in the signal round, and the announcement $\gamma$.
We denote the corresponding POVM as $\{\hat{E}^{\mathrm{obs}}_{\mathrm{bit},z,\gamma}\}_{z\in \{0,1\}, \gamma\in\cal W}\cup \{\hat{E}^{\mathrm{obs}}_{\mathrm{test},\gamma}\}_{\gamma \in\cal W_{\mathrm{test}}}$ in $\mathcal{L}(\mathcal{H}_A\otimes\mathcal{H}_B)$, i.e., the linear operators in $\mathcal{H}_A\otimes\mathcal{H}_B$.
From the protocol description, they can be expressed as
\begin{equation}
\begin{aligned}
 \hat{E}^{\mathrm{obs}}_{\mathrm{bit},z,\gamma} &= \sum_{\substack{(i,y,b,\alpha) \in \gamma: \\ f_{\mathrm{rec}}(i,y,b,\alpha)= z} } p_{\mathrm{aux},\alpha} \ket{i}\bra{i}_{A}\otimes \tau_B^y\Gamma^{b}_{y}, \quad \gamma \in \cal W \\
\hat{E}^{\mathrm{obs}}_{\mathrm{test},\gamma} & = p_{\mathrm{aux},\alpha} \ket{i}\bra{i}_{A}\otimes \tau_B^y\Gamma^{b}_{y}, \quad \gamma = \{(i,y,b,\alpha)\} \in \cal W_{\mathrm{test}}.
\end{aligned}
\end{equation}
We introduce a qubit system $\mathcal{H}_Q$, which is spanned by a set of orthogonal elements $\{\ket{0},\ket{1}\}$.
It can also be spanned by $\{\ket{+},\ket{-}\}$, where $\ket{\pm}$ is defined as $(\ket{0}\pm\ket{1})/\sqrt{2}$.
We choose a completely-positive trace-preserving (CPTP) map $\sum_{\gamma}\zeta_{\gamma}$ (each $\zeta_{\gamma}$ is a completely-positive map) from $\mathcal{H}_A\otimes\mathcal{H}_B$ to $\mathcal{H}_Q$ satisfying
\begin{equation}
 \label{eq:zeta-def}
 \begin{aligned}
 \hat{E}^{\mathrm{obs}}_{\mathrm{bit},z,\gamma} &=  \zeta^\dagger_{\gamma}\left(\ket{z}\bra{z}_{Q}\right),\quad z\in \{0,1\}, \gamma \in \cal W \\
 \zeta_{\gamma}(\hat{\rho}) & = \tr\left(\hat{E}^{\mathrm{obs}}_{\mathrm{test},\gamma}\hat{\rho}\right) \ket{\mathrm{0}}\bra{\mathrm{0}}_Q, \quad \hat{\rho} \in\mathcal{L}(\mathcal{H}_A\otimes\mathcal{H}_B), \gamma \in \cal W_{\mathrm{test}}.
 \end{aligned}
\end{equation}
The set $\{\zeta_\gamma\}_{\gamma}$ corresponds to a physical process where for the input state $\hat{\rho}$, the classical output $\gamma$ and the qubit state $\zeta_{\gamma}(\hat{\rho})/\tr(\zeta_{\gamma}(\hat{\rho}))$ is obtained with the probability $\tr(\zeta_{\gamma}(\hat{\rho}))$.
This physical process is called a completely-positive (CP) instrument.
Here $\zeta^\dagger_{\gamma}$ is the adjoint map of $\zeta_{\gamma}$, which satisfies $\tr(\zeta^\dagger_{\gamma}(\hat{O}_{Q})\hat{\rho}) =\tr(\hat{O}_{Q}\zeta_{\gamma}(\hat{\rho}))$ for any $\hat{O}_{Q}\in \mathcal{L}(\mathcal{H}_Q)$ and $\hat{\rho}\in\mathcal{L}(\mathcal{H}_A\otimes\mathcal{H}_B)$.
Then, the operator $\zeta^\dagger_{\gamma}\left(\ket{z}\bra{z}_{Q}\right)$ corresponds to an outcome that
$\gamma$ is obtained from the CP instrument $\{\zeta_\gamma\}_\gamma$ and $z$ is obtained from a $\{\ket{z}\bra{z}\}_{z\in\{0,1\}}$ measurement following the CP instrument.
We define a phase error operator $\hat{E}^{\mathrm{obs}}_{\mathrm{ph}}$ as 
\begin{equation}\label{eq:ephdef}
 \hat{E}^{\mathrm{obs}}_{\mathrm{ph}} = \sum_{\gamma\in \cal W}\hat{E}^{\mathrm{obs}}_{\mathrm{ph},-,\gamma},
\end{equation}
where $E^{\mathrm{obs}}_{\mathrm{ph},x,\gamma}$ is defined as
\begin{equation}\label{eq:ephgammadef}
 \hat{E}^{\mathrm{obs}}_{\mathrm{ph},x,\gamma} =  \zeta^\dagger_{\gamma}\left(\ket{x}\bra{x}_{Q}\right),\; x\in \{+,-\}, \gamma \in \cal W.
\end{equation}
Since the set of CP instrument $\{\zeta_\gamma\}_{\gamma}$ is not unique, the phase error operator $\hat{E}^{\mathrm{obs}}_{\mathrm{ph}}$ depends on the choice of $\{\zeta_\gamma\}_{\gamma}$.

Now we begin the asymptotic analysis based on phase error correction. The phase error ratio $e_{\mathrm{ph}}$, i.e., the ratio of the number of phase errors to the number of signal rounds, is a key parameter that characterizes the information leakage to the eavesdropper \cite{koashi2009simple,hayashi2012concise}.
In the following analysis, we assume the phase error operator $\hat{E}_{\mathrm{ph}}^{\mathrm{obs}}$ is given.
The value $\mathrm{tr}\left(\hat{E}_{\mathrm{ph}}^{\mathrm{obs}}\hat{\rho}_{\mathrm{out}}\right)$ characterizes the probability that the round is labeled as a signal round and the phase error occurs, we call it phase error rate per round here and hence force. We consider an asymptotic limit in which 
the quantity $N^{i,y,b}_{\mathrm{test}}/(p_{\mathrm{test}}(i,y,b)N_{\mathrm{tot}}\tau_A^i \tau_B^y)$ for each set of $(i,y,b)$ with $p_{\mathrm{test}}(i,y,b)>0$ converges to a value $q_{b|i,y}$ and $N_{\mathrm{sig}}/N_{\mathrm{tot}}$ converges to $Q_{\mathrm{sig}}$. The statistics $q_{b|i,y}$
give restrictions on $\hat{\rho}_{\mathrm{out}}$, $\mathrm{tr}( \hat{\rho}_{\mathrm{out}}  \ket{i}\bra{i} \otimes \Gamma_y^b) = \tau_A^i q_{b|i,y}$, so that we have a bound
\begin{equation}
\label{eq:maxeph}
\mathrm{tr}\left(\hat{E}_{\mathrm{ph}}^{\mathrm{obs}} \hat{\rho}_{\mathrm{out}}\right)
\leq Q_{\mathrm{sig}} e_{\mathrm{ph}}^U(\{q_{b|i,y}\}_{b,i,y}).
\end{equation}
Here $e_{\mathrm{ph}}^U(\{q_{b|i,y}\}_{b,i,y})$ is the upper bound of the phase error rate.
The key rate formula is then given in terms of $e_{\mathrm{ph}}^U(\{q_{b|i,y}\}_{b,i,y})$, 
\begin{equation}
\label{eq:general-keyrate-asymptotic}
R = Q_{\mathrm{sig}}(1-h(e_{\mathrm{ph}}^U(\{q_{b|i,y}\}_{b,i,y}))-f_{\mathrm{EC}}h(e_{\mathrm{bit}})),
\end{equation}
where $h(\cdot)$ denotes the binary Shannon entropy, $f_{\mathrm{EC}}$ is error correction efficiency, and $e_{\mathrm{bit}}$ is bit error rate. 
One way to find $e_{\mathrm{ph}}^U$ satisfying Eq.~\eqref{eq:maxeph} is to solve the following maximization problem for the phase error rate per round, 
\begin{equation}\label{eq:constraints}
\begin{aligned}
& \max_{\hat{\rho}}  \mathrm{tr}\left(\hat{E}_{\mathrm{ph}}^{\mathrm{obs}}\hat{\rho}\right) \\
\mathrm{s.t.} \quad & \hat{\rho} \succeq 0 \\
&  \mathrm{tr}_B(\hat{\rho})  =\mathrm{tr}_{A^\prime}(\hat{\rho}_{\mathrm{in}})\\
 & \mathrm{tr}( \hat{\rho}  \ket{i}\bra{i} \otimes \Gamma_y^b) = \tau_A^i q_{b|i,y},
\end{aligned}
\end{equation}
where the second constraint is due to the fact that the channel only acts on system $A^\prime$ of $\hat{\rho}_{\mathrm{in}}$ and keeps $A$ invariant. The third constraint means $\hat{\rho}$
should be compatible with the results in the test mode.


In the asymptotic analysis for some protocols, the optimal value of $\tau_A^i$ maximizing Eq.~\eqref{eq:general-keyrate-asymptotic} can approach zero for some $i$. This happens, for example, if the $i$-th state $\ket{\psi_i}$ is only prepared in the test rounds and we set $\tau_A^i = 0$, the formulation of Eq.~\eqref{eq:constraints} cannot handle the observed data associated with the $i$-th state properly.
To deal with this issue, we renormalize the density matrix $\hat{\rho}$ by defining the following invertible maps acting on an arbitrary operator $\hat{O}_A\in \mathcal{L}(\mathcal{H}_{A})$,
\begin{equation}
\begin{aligned}
\bra{i}\mathcal{T}_{\mathrm{PM}}(\hat{O}_A)\ket{i^\prime} & = 
\frac{1}{\sqrt{\tau_A^i \tau_A^{i^\prime}}}\bra{i^\prime}\hat{O}_A\ket{i} \\
\bra{i}\mathcal{T}^{-1}_{\mathrm{PM}}(\hat{O}_A)\ket{i^\prime} & = 
\sqrt{\tau_A^i \tau_A^{i^\prime}}\bra{i^\prime}\hat{O}_A\ket{i}, \\
\end{aligned}
\end{equation}
where the subscripts label the matrix elements. Then we have $\mathrm{tr}[\mathcal{T}^{-1}_{\mathrm{PM}}\otimes \mathcal{I}_B(\hat{O}_{AB})\hat{G}]= \mathrm{tr}[\hat{O}_{AB} \hat{\rho}]$ if $\hat{G} = \mathcal{T}_{\mathrm{PM}}\otimes \mathcal{I}_B(\hat{\rho})$ for an arbitrary operator $\hat{O}_{AB}\in \mathcal{L}(\mathcal{H}_{A}\otimes \mathcal{H}_{B})$. Here $\hat{G}$ is the renormalized density matrix. 

The renormalized phase error operator $\hat{E}_{\mathrm{ph}}$ is defined as
\begin{equation}
\hat{E}_{\mathrm{ph}} = \mathcal{T}^{-1}_{\mathrm{PM}}\otimes \mathcal{I}_B(\hat{E}_{\mathrm{ph}}^{\mathrm{obs}}). 
\end{equation}
Then the r.h.s of Eq.~\eqref{eq:maxeph} can be calculated by the following semi-definite programming (SDP) problem,
\begin{equation}
\label{eq:generalprimal}
\begin{aligned}
& \mathrm{tr}(\hat{E}_{\mathrm{ph}} \hat{G}) \\
\mathrm{s.t.} &  \quad \hat{G} \succeq 0 \\
 & \mathrm{tr}(\hat{Q}_{b,i|y}\hat{G}) = q_{b|i,y}\\
 & \mathrm{tr}(\hat{P}_{i,i^\prime}\otimes \hat{I}_B\hat{G}) = p_{i,i^\prime}, 
\end{aligned}
\end{equation}
where 
\begin{equation}\label{eq:operatordef}
\begin{aligned}
\hat{P}_{i,i^\prime} &= \left\{ 
\begin{array}{rcl}
\frac{1}{2}(\ket{i}\bra{i^\prime} + \ket{i^\prime}\bra{i}), \quad i \geq i^\prime  \\
\frac{1}{2i}(\ket{i}\bra{i^\prime} - \ket{i^\prime}\bra{i}),  \quad i< i^\prime\\
\end{array}
                           \right. \\
p_{i,i^\prime} &= \left\{ 
\begin{array}{rcl}
\mathrm{Re}(\braket{\psi_i}{\psi_{i^\prime}}), \quad i \geq i^\prime  \\
\mathrm{Im}(\braket{\psi_i}{\psi_{i^\prime}}),  \quad i< i^\prime\\
\end{array}
                           \right.         \\
\hat{Q}_{b,i|y} & = \ket{i}\bra{i} \otimes \Gamma_y^b.
\end{aligned}
\end{equation}


The constraints in Eq.~\eqref{eq:generalprimal} are classified into two categories. One is independent of Eve's attack, i.e., the inner product constraints $\mathrm{tr}(\hat{P}_{i,i^\prime}\otimes \hat{I}_B\hat{G}) = p_{i,i^\prime}$. Such inner product constraints come from the relation $\mathrm{tr}_{A^\prime}(\hat{\rho}_{\mathrm{in}})=\mathrm{tr}_{B}(\hat{\rho}_{\mathrm{out}})$; The other is determined by Eve's attack and compatible with the statistics observed in experiments, i.e., $\mathrm{tr}(\hat{Q}_{b,i|y}\hat{G}) = q_{b|i,y}$.

Though Eq.~\eqref{eq:generalprimal} seems to be independent from $\tau_A^i$ and $\tau_B^y$, according to the definition of phase error operator in Eq.~\eqref{eq:ephdef}, $\hat{E}_{\mathrm{ph}}^{\mathrm{obs}}$ and $\hat{E}_{\mathrm{ph}}$ depend on $\tau_A^i$ and $\tau_B^y$ in general. The dependence can be simple in some protocols, for example, $\hat{E}_{\mathrm{ph}}$ is proportional to $\tau_A^i$ and $\tau_B^y$. Then the optimization Eq.~\eqref{eq:generalprimal} only needs to be run once for different $\tau_A^i$ and $\tau_B^y$. One may refer to such an example in Sec.~\ref{Sec:PMQKD}.

Usually we do not make use of all experimental data or the inner product information. Such a case will be regarded as a relaxation of the constraints in Eq.~\eqref{eq:generalprimal}, which will be generally formulated as follows. We define new sets of operators and parameters as
\begin{equation}
\begin{aligned}
\hat{P}_k &= \sum_{i,i^\prime}\nu_{k,i,i^\prime} \hat{P}_{i,i^\prime} \\
p_k &= \sum_{i,i^\prime}\nu_{k,i,i^\prime} p_{i,i^\prime} \\
\hat{Q}_l &= \sum_{i,y,b:p_{\mathrm{test}}(i,y,b)>0} \beta_{l,i,y,b} \hat{Q}_{b,i|y} \\
q_l &= \sum_{i,y,b:p_{\mathrm{test}}(i,y,b)>0} \beta_{l,i,y,b} q_{b|i,y},
\end{aligned}
\end{equation}
where $\nu_{k,i,i^\prime}$ and $\beta_{l,i,y,b,\alpha}$ are real coefficients. The relaxed SDP is then given by 
\begin{equation}\label{eq:primalrelax}
\begin{aligned}
&\max_{\hat{G}} \mathrm{tr}(\hat{E}_{\mathrm{ph}}\hat{G}) \\
\mathrm{s.t.}  & \quad \hat{G} \succeq 0 \\
& \mathrm{tr}(\hat{P}_k\otimes \hat{I}_B\hat{G}) =p_k , \quad  k \in \{1,2,... m\}\\
& \mathrm{tr}(\hat{Q}_l\hat{G}) =q_l , \quad  l \in \{1,2,... n\}.\\
\end{aligned}
\end{equation}
the phase error calculated by Eq.~\eqref{eq:primalrelax} will never be less than Eq.~\eqref{eq:generalprimal}. If we define $N_l$ as

\begin{equation}\label{eq:nl}
N_l = \sum_{i,y,b:p_{\mathrm{test}}(i,y,b)>0} \frac{\beta_{l,i,y,b}N^{i,y,b}_{\mathrm{test}}}{\tau_A^i \tau_B^jp_{\mathrm{test}}(i,y,b)}.
\end{equation}
Then $N_l/N_{\mathrm{tot}}$ converges to $q_l$ in asymptotic limit.

\subsection{Finite-size analysis}\label{sec:finite}
Given an asymptotic analysis of a protocol, namely, a set of 
\begin{equation}
\nonumber
\{\ket{\psi_i}\}_i,\{\Gamma^b_y\}_{b,y}, f_{\mathrm{rec}}, \mathcal{W}, \{\hat{P}_k\}_k, \{\hat{Q}_l\}_l, \{\hat{E}^{\mathrm{obs}}_{\mathrm{ph},x,\gamma}\}_{x,\gamma},
\end{equation} 
we propose how to construct a protocol for which a secure key length is computable in the finite-key regime.
The protocol is parameterized by $\tau_A^i$, $\tau_B^y$, $p_{\mathrm{trash}}$, $p_{\mathrm{aux},\alpha}$, $q_l^{\mathrm{nom}}$, $\epsilon_{\mathrm{ph}}$, $\epsilon_{\mathrm{EC}}$, and $s_{\mathrm{PA}}$, which are called protocol parameters.
The description of the proposed protocol is as follows.

\textbf{Real Protocol}
\begin{enumerate}
	\item Alice randomly prepares a quantum state $\ket{\psi_i}\in \mathcal{H}_{A^\prime}$ with probability $\tau_A^i$ satisfying $\sum_i \tau_A^i =1$. Then she sends it to Bob through the channel. 
 Bob randomly chooses $y$ with probability $\tau_B^y$ satisfying $\sum_y \tau_B^y =1$
and conduct the generalized measurement specified by $\{\Gamma_y^b\}_b$, $\Gamma_y^b \in \mathcal{L}(\mathcal{H}_B)$ on the received signal.
Alice and Bob randomly choose auxiliary random variables for each round, which are collectively denoted as $\alpha$ with probability $p_{\mathrm{aux},\alpha}$. With probability $p_{\mathrm{trash}}$, Alice determines the round is a trash round.
 \item They repeat the process for $N_{\mathrm{tot}}$ rounds. Here, Alice can send all of her input states to the channel before Bob receives the states.
	\item Alice announces the position of the trash rounds among the $N_{\mathrm{tot}}$ rounds. Alice and Bob discard the recorded data $(i,y,b)$ for every trash round.
	\item For each round that is not a trash round, Alice and Bob announce $\gamma$ and determine the round is a signal round or a test round, corresponding to $\gamma\in \cal W$ and $\gamma \notin \cal W$, respectively. From the signal rounds, they generate the sifted key $z$ and record $N_{\mathrm{sig}}$ which is the number of signal rounds.
From the test rounds, they determine $N_l$ given by
\begin{equation}
N_l = \sum_{i,y,b:p_{\mathrm{test}}(i,y,b)>0} \frac{\beta_{l,i,y,b}N^{i,y,b}_{\mathrm{test}}}{\tau_A^i \tau_B^jp_{\mathrm{test}}(i,y,b)},
\nonumber
\end{equation}
where $N_{\mathrm{test}}^{i,y,b}$ is the number of test rounds with $(i,y,b)$.
	\item Alice and Bob perform the post-processing to generate a key of a length
	\begin{equation}
	 \label{eq:finite-keyrate-formula}
	K = N_{\mathrm{sig}}\left[1-h\left(\frac{M_{\mathrm{ph}}^U(N_{\mathrm{sig}},\{N_l\}_l,\epsilon_{\mathrm{ph}})}{N_{\mathrm{sig}}}\right)\right] -s_{\mathrm{PA}},
	\end{equation}
where $N_{\mathrm{sig}}$ is the number of signal rounds and $M_{\mathrm{ph}}^U(N_{\mathrm{sig}},\{N_l\}_l,\epsilon_{\mathrm{ph}})$ is a function determined by protocol parameters $q_l^{\mathrm{nom}}$.
Then the net key gain is $K - H_{\mathrm{EC}}$, where $H_{\mathrm{EC}}$ is the consumption of the pre-shared key in error correction process with a failure probability of $\epsilon_{\mathrm{EC}}$.
\end{enumerate}

To prove the finite-size security, we define a phase-control protocol as follows.

\textbf{Phase-control Protocol}
\begin{enumerate}
 \item  Alice prepares $\hat{\rho}_{\mathrm{in}}^{\otimes N_{\mathrm{tot}}}$ and sends system $\mathcal{H}_{A^\prime}^{\otimes N_{\mathrm{tot}}}$ to Bob through the channel. With probability $p_{\mathrm{trash}}$, Alice determines whether each round is a trash round. For each round that is not a trash round, Alice and Bob perform the CP instrument $\{\zeta_{\gamma}\}_\gamma$ and output a qubit.
	\item  Alice and Bob announce the the positions of the trash rounds. In the rounds not designated as trash rounds, they announce $\gamma$, record $N_{\mathrm{sig}}$, and determine $\{N_l\}_l$. For every test round, they discard the output qubit.
	\item  Alice and Bob output the $N_{\mathrm{sig}}$ qubits.
\end{enumerate}
From this definition, the phase-control protocol followed by a $\{\ket{0},\ket{1}\}$ measurement on $\mathcal{H}_Q$ amounts to carrying out the POVM,
\begin{equation}
\begin{aligned}
&\{(1-p_{\mathrm{trash}})\hat{E}^{\mathrm{obs}}_{\mathrm{bit},z,\gamma}\}_{z \in \{0,1\},\gamma\in\cal W} \\
& \cup\{(1-p_{\mathrm{trash}})\hat{E}^{\mathrm{obs}}_{\mathrm{test},\gamma}\}_{\gamma \in \cal {W}_{\mathrm{test}}} \cup \{p_{\mathrm{trash}}\hat{I}_{AB}\}.
\end{aligned}
\end{equation}
This means that the phase-control protocol followed by a $\{\ket{0},\ket{1}\}$ measurement on $\mathcal{H}_Q$ is equivalent to steps 1 to 4 in the real protocol, i.e., the classical-quantum state of the reconciled key system (after POVM) and Eve's system are the same. Therefore, we can analyze the secrecy of the sifted key bit $z$ by the phase-control protocol.

The phase-control protocol followed by a $\{\ket{+},\ket{-}\}$ measurement on $\mathcal{H}_Q$
amounts to that Alice and Bob performs a measurement
\begin{equation}
\begin{aligned}
& \{(1-p_{\mathrm{trash}})\hat{E}^{\mathrm{obs}}_{\mathrm{ph},x,\gamma}\}_{x \in \{+,-\},\gamma\in\cal W} \\
& \cup\{(1-p_{\mathrm{trash}})\hat{E}^{\mathrm{obs}}_{\mathrm{test},\gamma}\}_{\gamma \in \cal {W}_{\mathrm{test}}} \cup \{p_{\mathrm{trash}}\hat{I}_{AB}\}.
\end{aligned}
\end{equation}
on each output of the channel.
We denote $M_{\mathrm{ph}}$ as the number of phase error that corresponds to the POVM element $ (1-p_{\mathrm{trash}})\hat{E}^{\mathrm{obs}}_{\mathrm{ph},-,\gamma}$ for any $\gamma \in\cal W$.
If $M_{\mathrm{ph}}$ were zero, the reconciled sifted-key bit $z$ could be considered as a result of $\{\ket{0}\bra{0},\ket{1}\bra{1}\}$ measurement on the pure state $\ket{+}$. 
Since a pure state has no correlation with any system, it guarantees the secrecy.
In the case of non-zero $M_{\mathrm{ph}}$, we use a privacy amplification.
It is known that the privacy amplification corresponds to the correction of the phase error in the phase-control protocol.
It means that the final key can also be considered as a result of $\{\ket{0}\bra{0},\ket{1}\bra{1}\}$ measurement on the pure state $\ket{+}$.
This is a naive understanding about why the phase error correction approach can prove the security.

More precisely, suppose that there exists a function $M_{\mathrm{ph}}^U(N_{\mathrm{sig}},\{N_l\}_l,\epsilon_{\mathrm{ph}})$,
where $\epsilon_{\mathrm{ph}}$ is a real value satisfying $0 \leq \epsilon_{\mathrm{ph}} \leq 1$.
We assume that the function $M_{\mathrm{ph}}^U(N_{\mathrm{sig}},\{N_l\}_l,\epsilon_{\mathrm{ph}})$ satisfies
\begin{equation}
\label{eq:phaseerrorub}
\Pr(M_{\mathrm{ph}} > M_{\mathrm{ph}}^U(N_{\mathrm{sig}},\{N_l\}_l,\epsilon_{\mathrm{ph}}) )\leq \epsilon_{\mathrm{ph}}.
\end{equation}
The random variables $N_{\mathrm{sig}}$ and $\{N_l\}_l$ can be considered as the outcome of the phase-control protocol followed by a $\{\ket{+},\ket{-}\}$ measurement.
It means that $M_{\mathrm{ph}}^U(N_{\mathrm{sig}},\{N_l\}_l,\epsilon_{\mathrm{ph}})$ is an upper bound of $M_{\mathrm{ph}}$ which holds with a probability no smaller than $1-\epsilon_{\mathrm{ph}}$.
If we can find such a $M_{\mathrm{ph}}^U(N_{\mathrm{sig}},\{N_l\}_l,\epsilon_{\mathrm{ph}})$ and use it as $M_{\mathrm{ph}}^U$ in Eq.~\eqref{eq:finite-keyrate-formula}, there exists a classical post-processing \cite{koashi2009simple,hayashi2012concise,matsuura2020finite} guaranteeing that the protocol becomes $(\epsilon_{\mathrm{tot}}=\epsilon_{\mathrm{EC}}+\epsilon_{\mathrm{PA}})$-secure, where 
	\begin{equation}
	 \label{eq:finite-keyrate-e-secrecy}
	\epsilon_{\mathrm{PA}} =\sqrt{2(\epsilon_{\mathrm{ph}}+2^{-s_{\mathrm{PA}}})}.
	\end{equation}

The remaining problem is to obtain $M_{\mathrm{ph}}^U(N_{\mathrm{sig}},\{N_l\}_l,\epsilon_{\mathrm{ph}})$.
We introduce an estimation protocol where the joint probability distribution of $M_{\mathrm{ph}}$ and $\{N_l\}_l$ is the same as those in the phase-control protocol followed by a $\{\ket{+},\ket{-}\}$ measurement.
We assume that there is an observable $\hat{P}^{\mathrm{obs}}\in\mathcal{L}(\mathcal{H}_A)$, which will be defined later.
We denote its eigenstate and the corresponding eigenvalue as $\{(\ket{\omega_{w}},\omega_{w})\}_{w}$.
We define the estimation protocol as follows, which generates random variables
$\chi_{\mathrm{ph}}^{(u)}$, $\chi_{\mathrm{sig}}^{(u)}$, $\chi_{Q,l}^{(u)}$, and $\chi_{P}^{(u)}$ in the $u$-th round ($u = 1, 2, \cdots, N_{\mathrm{tot}}$).

\textbf{Estimation Protocol}
\begin{enumerate}[1']
\item Alice prepares $\hat{\rho}_{\mathrm{in}}$ and sends system $A^\prime$ to Bob through the channel. They perform the measurement whose POVM can be written as
				\begin{equation}
				 \label{eq:povmestimationprotocol2}
				 \begin{aligned}
				&  \{(1-p_{\mathrm{trash}}) E^{\mathrm{obs}}_{\mathrm{ph},x,\gamma}\}_{x\in\{+,-\},\gamma \in \cal W} \\
				 &\cup
                   \{(1-p_{\mathrm{trash}})\hat{E}^{\mathrm{obs}}_{\mathrm{test},\gamma}\}_{\gamma \in \cal{W}_{\mathrm{test}}} \cup
					\{p_{\mathrm{trash}}\ket{\omega_w}\bra{\omega_w}\otimes\hat{I}_{B}\}_w,
					\end{aligned}
				\end{equation}
				 to the output the channel in $\mathcal{L}(\mathcal{H}_A\otimes\mathcal{H}_B)$.



\begin{table}[hbt]
\begin{tabular}{c|c}
\hline
$\chi^{(u)}_{\mathrm{ph}}$ & POVM elements in Eq.~\eqref{eq:povmestimationprotocol2} \\
\hline
$1$ & $(1-p_{\mathrm{trash}})\hat{E}_{\mathrm{ph,-,\gamma}}^{\mathrm{obs}}$ \\ \hline
$0$  & other elements\\
\hline
\hline
$\chi^{(u)}_{\mathrm{sig}}$ & POVM elements in Eq.~\eqref{eq:povmestimationprotocol2} \\
\hline
 $1$ & $(1-p_{\mathrm{trash}})\hat{E}_{\mathrm{ph},x,\gamma}^{\mathrm{obs}}$, $x\in\{+,-\}$ \\ \hline
$0$  & other elements\\
\hline
\hline
$\chi^{(u)}_{Q,l}$ & POVM elements in Eq.~\eqref{eq:povmestimationprotocol2} \\
\hline
$\frac{\beta_{l,i,y,b}}{\tau_A^i\tau_B^y p_{\mathrm{test}}(i,y,b)} $  
 & $(1-p_{\mathrm{trash}})\hat{E}_{\mathrm{test},\gamma}^{\mathrm{obs}}$, $\gamma = \{(i,y,b,\alpha)\}$  \\ \hline
$0$  & other  elements\\
\hline
\hline
$\chi^{(u)}_{P}$ & POVM elements in Eq.~\eqref{eq:povmestimationprotocol2} \\
\hline
 $\omega_w$ & $p_{\mathrm{trash}}\ket{\omega_w}\bra{\omega_w}\otimes\hat{I}_{B}$\\ \hline
$0$  & other elements\\
\hline
\end{tabular}
\caption{The correspondence between the values of the random variables and the POVM elements.}
\label{tab:randomvariablechi}
\end{table}
\item Alice and Bob obtain the values of random variables $\chi_{\mathrm{ph}}^{(u)}$, $\chi_{\mathrm{sig}}^{(u)}$, $\chi_{Q,l}^{(u)}$, and $\chi_{P}^{(u)}$ in each round. The correspondence between the values of the random variables and the POVM elements is given in Table~\ref{tab:randomvariablechi}. 
\item They repeat steps 1 and 2 for $N_{\mathrm{tot}}$ rounds and record $M_{\mathrm{ph}}=\sum_{u=1}^{N_{\mathrm{tot}}} \chi_{\mathrm{ph}}^{(u)}$, $N_{\mathrm{sig}}=\sum_{u=1}^{N_{\mathrm{tot}}} \chi_{\mathrm{sig}}^{(u)}$ and $N_l = \sum_{u=1}^{N_{\mathrm{tot}}} \chi_{Q,l}^{(u)}$.
\end{enumerate}

In the phase-control protocol, Alice and Bob just discard their data in the trash rounds.
It means that the the joint probability distributions of random variables $\chi_{\mathrm{ph}}^{(u)}$, $\chi_{Q,l}^{(u)}$, 
and the positions of the trash rounds in the phase-control protocol and the estimation protocol are the same.
The protocol is proved to be secure if we can show Eq.~\eqref{eq:phaseerrorub}.


In order to analyze the statistics in the estimation protocol, we will use a linear operator inequality in the following form
\begin{equation}\label{eq:opineq}
\hat{E}_{\mathrm{ph}} +\sum_k \lambda_k \hat{P}_k \otimes \hat{I}_B + \sum_l \eta_l \hat{Q}_l   \preceq 0,
\end{equation}
to apply the concentration inequalities. To obtain a suitable operator inequality, we propose to consider an alternative method other than Eq.~\eqref{eq:primalrelax} for asymptotic analysis, which can give rise to a linear operator inequality in the form of Eq.~\eqref{eq:opineq} and is easy to be extended into finite-size analysis. The alternative method works as follows. 
We first estimate $\vec{q}=(q_1, q_2, \cdots, q_l, \cdots, q_n)$ in Eq.~\eqref{eq:primalrelax} with the protocol parameters $q_l^{\mathrm{nom}}$ prior to the actual run of the protocol. 
The estimations $\vec{q}^\mathrm{nom}=(q_1^\mathrm{nom}, q_2^\mathrm{nom}, \cdots, q_l^{\mathrm{nom}}, \cdots, q_n^\mathrm{nom})$ are called nominal values. Then we consider the dual problem of Eq.~\eqref{eq:primalrelax},
\begin{equation}\label{eq:dual}
\begin{aligned}
& \min_{\vec{\Lambda}} - \vec{\Lambda} \cdot \vec{C} \\
\mathrm{s.t.} \quad & \hat{E}_{\mathrm{ph}} + \sum_k \lambda_k \hat{P}_k \otimes \hat{I}_B + \sum_l \eta_l \hat{Q}_l \preceq 0.
\end{aligned}
\end{equation}
where $\vec{\Lambda}=(\lambda_1, \lambda_2, \cdots, \lambda_m, \eta_1, \eta_2, \cdots, \eta_n)$, $\vec{C}=(\vec{p}, \vec{q}^\mathrm{nom})$ and $\vec{p}=(p_1, p_2, \cdots, p_m)$. Suppose the optimal $\vec{\Lambda}$ for Eq.~\eqref{eq:dual} is $\vec{\Lambda}^*(\vec{q}^\mathrm{nom})$. After the actual run of the protocol, we can observe true values of $\vec{q}$, denoted as $\vec{q}^\mathrm{obs}=(q_1^\mathrm{obs}, q_2^\mathrm{obs}, \cdots, q_n^\mathrm{obs})$, with which we can calculate an upper bound of the phase error rate $p^*_\mathrm{lin}(\vec{q}^\mathrm{obs};\vec{q}^\mathrm{nom}) = -\vec{\Lambda}^*(\vec{q}^\mathrm{nom}) \cdot (\vec{p},\vec{q}^\mathrm{obs})$. The alternative method uses $p^*_\mathrm{lin}(\vec{q}^\mathrm{obs};\vec{q}^\mathrm{nom})$ as the result, which is different from the optimal value of the primal problem Eq.~\eqref{eq:primalrelax}, $p^*(\vec{q}^\mathrm{obs})$.

Since $\vec{\Lambda}^*(\vec{q}^\mathrm{nom})$ satisfies the constraint in Eq.~\eqref{eq:dual}, we have the following inequality,
\begin{equation}\label{eq:opineqopt}
\hat{E}_{\mathrm{ph}} + \sum_k \lambda_k^* \hat{P}_k \otimes \hat{I}_B + \sum_l \eta^*_l \hat{Q}_l\preceq 0,
\end{equation}
which has the same form as Eq.~\eqref{eq:opineq}. With the inequality Eq.~\eqref{eq:opineqopt}, we can obtain $\mathrm{tr}(\hat{E}_{\mathrm{ph}}\hat{G}) \leq  p^*_\mathrm{lin}(\vec{q}^\mathrm{obs};\vec{q}^\mathrm{nom})$ for a renormalized density matrix $\hat{G}$ satisfying the constraints in Eq.~\eqref{eq:primalrelax}, which means the alternative method can give a valid upper bound on the phase error rate. Therefore, the alternative method is secure.

We compare the alternative method with the original one in Eq.~\eqref{eq:primalrelax}. Their performances on asymptotic key rate depend on $p^*_\mathrm{lin}(\vec{q}^\mathrm{obs};\vec{q}^\mathrm{nom})$ and $p^*(\vec{q}^\mathrm{obs})$, respectively. According to the Lagrange duality of Eq.~\eqref{eq:primalrelax} and Eq.~\eqref{eq:dual}, we have $-\vec{\Lambda}^*(\vec{q}^\mathrm{obs}) \cdot (\vec{p},\vec{q}^\mathrm{obs}) \geq p^*(\vec{q}^\mathrm{obs})$, which saturates when the strong duality holds. Actually, the strong duality holds in most practical cases, unless the observed statistics $\vec{q}^\mathrm{obs}$ forbid the possibility of at least one pure state, in other words, there is no full-rank solution in Eq.~\eqref{eq:primalrelax}.
If we take $\vec{C}=(\vec{p}, \vec{q}^\mathrm{obs})$ in Eq.~\eqref{eq:dual}, then we have $-\vec{\Lambda}^*(\vec{q}^\mathrm{nom}) \cdot (\vec{p},\vec{q}^\mathrm{obs}) \geq -\vec{\Lambda}^*(\vec{q}^\mathrm{obs}) \cdot (\vec{p},\vec{q}^\mathrm{obs})$ since $-\vec{\Lambda}^*(\vec{q}^\mathrm{nom})$ satisfies the constraint in Eq.~\eqref{eq:dual} and $\vec{\Lambda}^*(\vec{q}^\mathrm{obs})$ minimizes $-\vec{\Lambda} \cdot \vec{C}$ over all $\vec{\Lambda}$ satisfying that constraint. This inequality saturates when $\vec{q}^\mathrm{obs}=\vec{q}^\mathrm{nom}$. Combining the two inequalities above, we have $p^*_\mathrm{lin}(\vec{q}^\mathrm{obs};\vec{q}^\mathrm{nom})\geq p^*(\vec{q}^\mathrm{obs})$. Therefore, the asymptotic key rate given by the alternative method is in general worse than the original one. The two key rates coincide only when the strong duality holds and $\vec{q}^\mathrm{obs}=\vec{q}^\mathrm{nom}$.

Here we make some remarks on the alternative method. The elements of $\vec{q}^\mathrm{nom}$ are fixed prior to the actual run of the protocol so that they are independent of the observed experimental data. For example, one may compute $\vec{q}^\mathrm{nom}$ by assuming a channel model, or carry out test rounds to estimate $\vec{q}^\mathrm{nom}$. This is different from the literatures of asymptotic numerical frameworks \cite{coles2016numerical, winick2018reliable, wang2019characterising,primaatmaja2019versatile} where the observed experiment results can be directly substituted into the optimization problems.
We use the nominal values for the ease of applying concentration inequalities.

Now we extend the alternative method to finite-size analysis. We define $\hat{P}^\mathrm{obs}= \mathcal{T}_{\mathrm{PM}}(\sum_k \lambda_k^* \hat{P}_k)=\sum_{\omega_w} \omega_w \ket{\omega_w}\bra{\omega_w}$.
Then the operator inequality given by the alternative protocol Eq.~\eqref{eq:opineqopt} can be rewritten as $\hat{E}_{\mathrm{ph}} + \mathcal{T}_{\mathrm{PM}}^{-1}(\hat{P}^\mathrm{obs}) \otimes I_B + \sum_l \eta^*_l \hat{Q}_l\preceq 0$. We can analyze the statistics in the estimation protocol by combining this operator inequality and concentration inequalities.

The conditional expectation values of random variables $\chi_{\mathrm{ph}}^{(u)}$, $\chi_{Q,l}^{(u)}$, and $\chi_P^{(u)}$ are given by
\begin{equation}\label{eq:conditionalexpec}
\begin{aligned}
E(\chi_{\mathrm{ph}}^{(u)}|F^{u-1}) 
& =\mathrm{tr}\left[(1-p_{\mathrm{trash}})\sum_{\gamma\in \cal W}\hat{E}_{\mathrm{ph,-,\gamma}}^{\mathrm{obs}} \hat{\rho}_{\mathrm{out}}^{F^{u-1}}\right]  \\
& = (1-p_{\mathrm{trash}}) \mathrm{tr}\left(\hat{E}_{\mathrm{ph}}^{\mathrm{obs}} \hat{\rho}_{\mathrm{out}}^{F^{u-1}}\right) \\
E(\chi_{Q,l}^{(u)}|F^{u-1}) 
&= \sum_{\gamma \in {\cal W}_{\mathrm{test}}}\frac{\beta_{l,i,y,b}}{\tau_A^i \tau_B^y p_{\mathrm{test}}(i,y,b)}  \\ 
& \phantom{=\;\;}\mathrm{tr}\left[(1-p_{\mathrm{trash}}) \hat{E}_{\mathrm{test},\gamma}^{\mathrm{obs}} \hat{\rho}_{\mathrm{out}}^{F^{u-1}}\right] \\
& = \sum_{\substack{i,y,b: \\p_{\mathrm{test}(i,y,b)>0}}}\frac{\beta_{l,i,y,b}}{\tau_A^i \tau_B^y p_{\mathrm{test}}(i,y,b)} \\ 
& \phantom{=\;\;} \mathrm{tr}\left[(1-p_{\mathrm{trash}}) \sum_{\alpha}\hat{E}_{\mathrm{test},\gamma}^{\mathrm{obs}} \hat{\rho}_{\mathrm{out}}^{F^{u-1}}\right] \\
& = (1-p_{\mathrm{trash}}) \mathrm{tr}\left[\mathcal{T}_{\mathrm{PM}}\otimes \mathcal{I}_B(\hat{Q}_l)\hat{\rho}_{\mathrm{out}}^{F^{u-1}}\right] \\
E(\chi_{P}^{(u)}|F^{u-1}) & = \sum_{\omega_x}\omega_x\mathrm{tr}\left[p_{\mathrm{trash}} \ket{\omega_x}\bra{\omega_x}\otimes \hat{I}_B \hat{\rho}_{\mathrm{out}}^{F^{u-1}}\right] \\
& = {p_{\mathrm{trash}}}\mathrm{tr}\left( \hat{P}^{\mathrm{obs}} \otimes \hat{I}_B\hat{\rho}_{\mathrm{out}}^{F^{u-1}}\right).
\end{aligned}
\end{equation}
where $F^{u-1}$ is the filtration identifying random variables $\chi_{\mathrm{ph}}^{(u^\prime)}$, $\chi_{Q,l}^{(u^\prime)}$, and $\chi_P^{(u^\prime)}$ ($l\in\{1,2,\cdots,n\}$, $u^\prime \in \{1,2,\cdots, u-1\}$) and $\hat{\rho}_{\mathrm{out}}^{F^{u-1}}$ is the post-measurement density matrix (the measurement results of the first $u-1$ rounds are the same as the filtration). Recalling Eq.~\eqref{eq:opineqopt}, we have the following inequality
\begin{equation}\label{eq:traceopineq}
\begin{aligned}
& \mathrm{tr}\left(\hat{E}_{\mathrm{ph}}^{\mathrm{obs}} \hat{\rho}_{\mathrm{out}}^{F^{u-1}}\right)+ \sum_{l=1}^n\mathrm{tr}\left[\mathcal{T}_{\mathrm{PM}}\otimes \mathcal{I}_B(\eta_l^*\hat{Q}_l)\hat{\rho}_{\mathrm{out}}^{F^{u-1}}\right] \\& +  \mathrm{tr}\left( \hat{P}^{\mathrm{obs}} \otimes \hat{I}_B\hat{\rho}_{\mathrm{out}}^{F^{u-1}}\right)\leq 0.
\end{aligned}
\end{equation} 
Combining Eq.~\eqref{eq:conditionalexpec} and Eq.~\eqref{eq:traceopineq}, we have the following inequality for conditional expectation values
\begin{equation}
\label{eq:ceineq}
\begin{aligned}
& \frac{1}{1-p_{\mathrm{trash}}}E(\chi_{\mathrm{ph}}^{(u)}|F^{u-1})+ \frac{1}{1-p_{\mathrm{trash}}}\sum_{l=1}^n\eta_l^* E(\chi_{Q,l}^{(u)}|F^{u-1}) \\& +\frac{1}{p_{\mathrm{trash}}}E(\chi_{P}^{(u)}|F^{u-1})\leq 0.
\end{aligned}
\end{equation}

To deal with the statistical fluctuations in $\chi_{\mathrm{ph}}^{(u)}$, $\chi_{Q,l}^{(u)}$ and $\chi_{P}^{(u)}$, we may apply concentration inequalities for dependent variables such as the Azuma's inequality or Kato's inequality \cite{kato2020concentration}.
Suppose we obtain an inequality in the following form,
\begin{equation}\label{eq:conineq12}
\begin{aligned}
& \sum_{u=1}^{N_{\mathrm{tot}}} \left[\left(\frac{\chi_{\mathrm{ph}}^{(u)}+\sum_{l=1}^n \eta_l^* \chi_{Q,l}^{(u)}}{1-p_{\mathrm{trash}}} + \frac{\chi_P^{(u)}}{p_{\mathrm{trash}}}\right) \right.\\ &\left.\phantom{=\;\;}- E\left(\frac{\chi_{\mathrm{ph}}^{(u)}+\sum_{l=1}^n \eta_l^* \chi_{Q,l}^{(u)}}{1-p_{\mathrm{trash}}} +\frac{\chi_P^{(u)}}{p_{\mathrm{trash}}}|F^{u-1}\right) \right]  \\& \leq \Delta_1(N_{\mathrm{sig}},\{N_l\}_l,N_{\mathrm{tot}},\frac{\epsilon_\mathrm{ph}}{2}),
\end{aligned}
\end{equation}
which holds with a probability no smaller than $1-\epsilon_\mathrm{ph}/2$. Note that one may apply a concentration inequality to a single variable 
$(\chi_{\mathrm{ph}}^{(u)}+\sum_{l=1}^n \eta^*_l \chi_{Q,l}^{(u)})/(1-p_{\mathrm{trash}}) +\chi_P^{(u)}/p_{\mathrm{trash}}$, or apply concentration inequalities separately to each variable and then combine the results with the union bound.
We note that the relation $\chi_{\mathrm{ph}}^{(u)}\leq \chi_{\mathrm{sig}}^{(u)}$, which is derived from their definitions in the Table~\ref{tab:randomvariablechi}, is useful when we apply Kato's inequality to $\chi_{\mathrm{ph}}^{(u)}$.

Recalling Eq.~\eqref{eq:operatordef} and the definition of $\hat{P}^{\mathrm{obs}}$, we can see that $\chi_{P}^{(u)}$ is independent and identically distributed (i.i.d.). We may apply concentration inequalities for independent variables such as the Chernoff-bound, and have an inequality in the following form
\begin{equation}\label{eq:conineq3}
-\Delta_2(N_{\mathrm{tot}},\frac{\epsilon_\mathrm{ph}}{2}) \leq \sum_{u=1}^{N_{\mathrm{tot}}} \chi_{P}^{(u)} -  N_{\mathrm{tot}} \sum_{k=1}^m \lambda_k^* p_k ,
\end{equation}
which holds with a probability no smaller than $1-\epsilon_\mathrm{ph}/2$. Combining Eqs.~\eqref{eq:ceineq}, \eqref{eq:conineq12} and \eqref{eq:conineq3}, we have
\begin{equation}\label{eq:ephub}
\begin{aligned}
\sum_{u=1}^{N_{\mathrm{tot}}}\chi_{\mathrm{ph}}^{(u)} &  \leq -\sum_{u=1}^{N_{\mathrm{tot}}}\sum_{l=1}^n \eta_l^* \chi_{Q,l}^{(u)}- \frac{1-p_{\mathrm{trash}}}{p_{\mathrm{trash}}}N_{\mathrm{tot}} \sum_{k=1}^m \lambda_k^* p_k \\ & \phantom{=\;\;}+ (1-p_{\mathrm{trash}}) \Delta_1(N_{\mathrm{sig}},\{N_l\}_l,N_{\mathrm{tot}},\frac{\epsilon_\mathrm{ph}}{2}) \\& \phantom{=\;\;} + \frac{1-p_{\mathrm{trash}}}{p_{\mathrm{trash}}}\Delta_2(N_{\mathrm{tot}},\frac{\epsilon_\mathrm{ph}}{2}) \\ &= M_{\mathrm{ph}}^U(N_{\mathrm{sig}},\{N_l\}_l,\epsilon_\mathrm{ph}),
\end{aligned}
\end{equation}
which holds with probability at least $1-\epsilon_{\mathrm{ph}}$. 
Therefore, our protocol is $(\epsilon_{\mathrm{tot}}=\epsilon_{\mathrm{EC}}+\epsilon_{\mathrm{PA}})$-secure.

In our protocol description, the classical communication between Alice and Bob begins only after the quantum communication is over. In addition, the number of total rounds is fixed prior to the quantum communication. One should be careful when applying our method to some protocols where Alice and Bob exchange the basis information in each round, such as those with iterative sifting \cite{pfister2016sifting}. In these protocols, some concentration inequalities such as the Chernoff-bound may fail due to the requirement of i.i.d. random variables.

Recently, there are also some researches on asymptotic numerical security analysis of QKD based on Gram matrix formulation \cite{wang2019characterising,primaatmaja2019versatile}. We compare our method with them for asymptotic analysis. The main difference lies in Bob's dimension. In our case Bob's dimension is determined by the application of squashing model or tagging method (see Appendix~\ref{app:tag}), while in \cite{wang2019characterising} the dimension depends on the number of POVM outcomes. In order to get better key rate, higher hierarchical structure is introduced to the Gram matrix, which will enlarge Bob's dimension. Another difference is that the Gram matrix method is semi-device independent, i.e., the analysis is independent of the implementation of measurement settings. In our work, the renormalized density matrix can also be regarded as a Gram matrix. Its elements are the inner products of the states left for Eve when Alice sends some certain state $\ket{\psi_i}$ and Bob has a measurement outcome on his basis $\{\ket{v}\}$. 

We also compare our numerical methods for asymptotic case and finite-size case. To calculate the asymptotic key rate, the constraints $\mathrm{tr}(\hat{Q}_l\hat{G}) =q_l$ should include as many experimental data as possible. However, this is not always true in finite-size case since more observables will introduce larger deviations due to the applications of concentration inequalities.

\section{Numerical framework for measurement-device-independent QKD protocols}\label{sec:mdi}
Now we consider a broad type of measurement-device-independent (MDI) QKD protocols among the communication partners, Alice and Bob, and an untrusted node in the middle, Charlie. Alice(Bob) randomly prepares a quantum state $\ket{\psi_i}$($\ket{\psi_j}$) with probability $\tau_A^i$($\tau_B^j$) satisfying $\sum_{i=1}^{d_A} \tau_A^i=1$($\sum_{j=1}^{d_B} \tau_B^j=1$). They send the signals to Charlie, who is supposed to jointly measure their signals and announce the results $\xi$. Similarly to prepare-and-measure protocols, Alice and Bob may use additional random values which are collectively represented by $\alpha$ with probability $p_{\mathrm{aux},\alpha}$.
Then they publicly communicate a part of the results $(i,j,\alpha)$ abstractly denoted as $\tilde{\gamma}$, which is represented by a map $f^\xi_{\mathrm{ann}}$ from $(i,j,\alpha)$ to $\tilde{\gamma}$. They determine the role of each round based on $\tilde{\gamma}$ and a predetermined set $W^\xi$ such that $\tilde{\gamma}\in W^\xi$ for a signal round and $\tilde{\gamma} \notin W^\xi$ for a test round.  In a signal round, they generate sifted key bit $z\in\{0,1\}$ represented by a function $f^\xi_{\mathrm{rec}}(i,j,\alpha)$. In a test round, they announce $(i,j,\alpha)$. They record the number of test rounds with $(i,j,\xi)$, $N_{\mathrm{test}}^{i,j,\xi}$, the number of signal rounds $N_{\mathrm{sig}}$. Finally they perform the post-processing and extract secure key. Similar to the prepare-and-measure protocols, we denote the whole announcement in each round as $\gamma$ and define the probability that $\alpha$ belongs to a test round as $p_{\mathrm{test}}(i,j,\xi) = \sum_{\alpha:f^\xi_{\mathrm{ann}}(i,j,\alpha)\notin W^\xi} p_{\mathrm{aux},\alpha}$. The sets $\mathcal{W}$, $\mathcal{W}_{\mathrm{test}}$, and $\mathcal{W}_{\mathrm{all}}$ are also defined for each value of $\xi$, which we denote by $\mathcal{W}^\xi$, $\mathcal{W}^\xi_{\mathrm{test}}$, and $\mathcal{W}^\xi_{\mathrm{all}}$.

In the corresponding entanglement-based protocol, Alice and Bob prepare entangled quantum states $\sum_{i=1}^{d_A} \sqrt{\tau_A^i}\ket{i}_{A} \ket{\psi_i}_{A^\prime}$ and $\sum_{j=1}^{d_B} \sqrt{\tau_B^j}\ket{j}_{B} \ket{\psi_j}_{B^\prime}$, respectively. Then the total state in $\mathcal{H}_{A}\otimes\mathcal{H}_{B}\otimes\mathcal{H}_{A^\prime}\otimes\mathcal{H}_{B^\prime}$ is given by
\begin{equation}\label{eq:mdidensity}
\hat{\rho}_{\mathrm{in}} = \sum_{i,i^\prime,j,j^\prime}\sqrt{\tau_A^i \tau_A^{i^\prime} \tau_B^j \tau_B^{j^\prime}}\ket{i}\bra{i^\prime}_A\otimes\ket{\psi_i}\bra{\psi_{i^\prime}}_{A^\prime}\otimes \ket{j}\bra{j^\prime}_{B}\otimes \ket{\psi_j}\bra{\psi_{j^\prime}}_{B^\prime}.
\end{equation} 
Alice and Bob keep their ancillary system in $\mathcal{H}_{A}\otimes\mathcal{H}_{B}$ and send the signal parts in $\mathcal{H}_{A^\prime}\otimes\mathcal{H}_{B^\prime}$ to Charlie. Then the dimension of $\mathcal{H}_{A}$ and $\mathcal{H}_{B}$ are $d_A$ and $d_B$, respectively. They repeat the process for $N_{\mathrm{tot}}$ rounds and share a bipartite state in $\mathcal{H}_{A}^{\otimes N_{\mathrm{tot}}} \otimes \mathcal{H}_{B}^{\otimes N_{\mathrm{tot}}}$.

We also represent Alice's and Bob's processes of each round in the actual protocol, which depend on Charlie's announcement $\xi$ in this case, as a POVM in the entanglement-based protocol. The POVM elements $E^{\mathrm{obs},\xi}_{\mathrm{bit},z,\gamma}$ and $E^{\mathrm{obs},\xi}_{\mathrm{test}, \gamma}$ satisfy
\begin{equation}\label{eq:povmmdi}
\begin{aligned}
 E^{\mathrm{obs},\xi}_{\mathrm{bit},z,\gamma} &= \sum_{\substack{(i,j,\alpha)\in \gamma:\\f^\xi_{\mathrm{rec}}(i,j,\alpha)= z}  } p_{\mathrm{aux},\alpha} \ket{i}\bra{i}_{A}\otimes \ket{j}\bra{j}_B, \quad \gamma \in \cal W^\xi\\
 E^{\mathrm{obs},\xi}_{\mathrm{test}, \gamma} &= p_{\mathrm{aux},\alpha} \ket{i}\bra{i}_{A}\otimes \ket{j}\bra{j}_B, \quad \gamma = \{(i,j,\alpha)\} \in \cal W^\xi_{\mathrm{test}}.
 \end{aligned}
\end{equation}
The qubit system $\mathcal{H}_Q$ and the CP instrument $\{\zeta_\gamma^\xi\}_\gamma$ are similarly defined as those in prepare-and-measure protocols.


The phase error rate is upper bounded by the following maximization problem,
\begin{equation}\label{eq:optmdi}
\begin{aligned}
&\max_{\hat{\rho}} \mathrm{tr}\left(\sum_\xi \hat{E}_{\mathrm{ph}}^{\mathrm{obs},\xi} \hat{\rho}^\xi\right)  \\
\mathrm{s.t.} \quad &\mathrm{tr}_{A^\prime B^\prime}(\hat{\rho}_{\mathrm{in}})  = \sum_\xi \hat{\rho}^\xi \\
 &\mathrm{tr}( \hat{\rho}^\xi  \ket{ij}\bra{ij}) = \tau_A^i \tau_B^j q_{\xi|i,j},
\end{aligned}
\end{equation}
where $q_{\xi|i,j}$ is the convergence of $N_{\mathrm{test}}^{i,j,\xi}/(p^\xi_{\mathrm{test}} N_{\mathrm{tot}}\tau_A^i \tau_B^j)$ in asymptotic limit. The phase error operator $\hat{E}_{\mathrm{ph}}^{\mathrm{obs},\xi}$ is defined following Eqs.~\eqref{eq:zeta-def} to \eqref{eq:ephgammadef} for each value of Charlie's announcement $\xi$. The key rate formula is given by
\begin{equation}
\label{eq:general-keyrate-asymptotic-mdi}
R = Q_{\mathrm{sig}}(1-h(e_{\mathrm{ph}}^U(\{q_{\xi|i,j}\}_{\xi,i,j}))-f_{\mathrm{EC}}h(e_{\mathrm{bit}})),
\end{equation}
where $Q_{\mathrm{sig}}$ is the convergence of $N_{\mathrm{sig}}/N_{\mathrm{tot}}$ in asymptotic limit and $e_{\mathrm{ph}}^U(\{q_{\xi|i,j}\}_{\xi,i,j})$ is obtained by solving Eq.~\eqref{eq:optmdi}.

For renormalization, we consider the following invertible maps for an arbitrary operator $\hat{O}_{AB}\in \mathcal{L}(\mathcal{H}_{A}\otimes \mathcal{H}_{B})$,
\begin{equation}
\begin{aligned}
\bra{ij}\mathcal{T}_{\mathrm{MDI}}(\hat{O}_{AB})\ket{i^\prime j^\prime} &= \frac{1}{\sqrt{\tau_A^i \tau_A^{i^\prime} \tau_B^j \tau_B^{j^\prime}} } \bra{i^\prime j^\prime}\hat{O}_{AB}\ket{i j} \\
\bra{ij}\mathcal{T}^{-1}_{\mathrm{MDI}}(\hat{O}_{AB})\ket{i^\prime j^\prime}  &= \sqrt{\tau_A^i \tau_A^{i^\prime} \tau_B^j \tau_B^{j^\prime}}  \bra{i^\prime j^\prime}\hat{O}_{AB}\ket{i j}.
\end{aligned}
\end{equation}
We define the renormalized density matrix and phase error operator $\hat{E}_{\mathrm{ph}}$ as
\begin{equation}
\begin{aligned}
& \hat{G}  = \sum_\xi \mathcal{T}_{\mathrm{MDI}} (\hat{\rho}^\xi)\otimes \ket{\xi}\bra{\xi}_\Xi \\
& \hat{E}_{\mathrm{ph}}= \sum_\xi \mathcal{T}^{-1}_{\mathrm{MDI}} (\hat{E}_{\mathrm{ph}}^{\mathrm{obs},\xi}) \otimes \ket{\xi}\bra{\xi}_\Xi,
\end{aligned}
\end{equation}
where we introduced the Hilbert space ${\cal H}_\Xi$ for Charlie's announcement $\xi$ for convenience.
Then we have 
\begin{equation}
\mathrm{tr}\left(\sum_\xi \hat{E}_{\mathrm{ph}}^{\mathrm{obs},\xi} \hat{\rho}^\xi\right) = \mathrm{tr}\left(\hat{E}_{\mathrm{ph}} \hat{G}\right).
\end{equation}
The upper bound of the phase error rate is calculated by the following SDP problem,
\begin{equation}\label{eq:generalprimalmdi}
\begin{aligned}
& \max_{\hat{G}} \mathrm{tr}(\hat{E}_{\mathrm{ph}} \hat{G}) \\
\mathrm{s.t.} &  \quad \hat{G} \succeq 0 \\
 & \mathrm{tr}(\hat{Q}_{\xi,i,j}\hat{G}) = q_{\xi|i,j} \\
 & \mathrm{tr}(\hat{P}_{i,i^\prime,j,j^\prime}\otimes \hat{I}_{\Xi}\hat{G}) = p_{i,i^\prime,j,j^\prime} \\
 & \mathrm{tr}(\hat{I}_{AB} \otimes \ket{\xi}\bra{\xi^\prime}_\Xi\hat{G}) = 0, \quad \xi \neq \xi^\prime,
\end{aligned}
\end{equation}
where 
\begin{equation}
\begin{aligned}
\hat{P}_{i,i^\prime,j,j^\prime} &= \left\{ 
\begin{array}{rcl}
\frac{1}{2}(\ket{ij}\bra{i^\prime j^\prime} + \ket{i^\prime j^\prime}\bra{i j}), \quad id_B+j\geq i^\prime d_B + j^\prime \\
\frac{1}{2i}(\ket{ij}\bra{i^\prime j^\prime} - \ket{i^\prime j^\prime}\bra{ij}),\quad id_B+j< i^\prime d_B + j^\prime\\
\end{array}
                           \right. \\
p_{i,i^\prime,j,j^\prime} &= \left\{ 
\begin{array}{rcl}
\mathrm{Re}(\braket{\psi_i}{\psi_{i^\prime}}\braket{\psi_j}{\psi_{j^\prime}}), \quad id_B+j\geq i^\prime d_B + j^\prime \\
\mathrm{Im}(\braket{\psi_i}{\psi_{i^\prime}}\braket{\psi_j}{\psi_{j^\prime}}),  \quad id_B+j< i^\prime d_B + j^\prime \\
\end{array}
                           \right.         \\
\hat{Q}_{\xi,i,j} & = \ket{i,j}\bra{i,j} \otimes \ket{\xi}\bra{\xi}_\Xi.
\end{aligned}
\end{equation}
The last constraint in Eq.~\eqref{eq:generalprimalmdi} comes from the fact that the register for Charlie's announcement is classical. 

In practical cases we also consider the relaxation of the constraints,
\begin{equation}\label{eq:relaxationMDI}
\begin{aligned}
\hat{P}_k &= \sum_{i,i^\prime,j,j^\prime}\nu_{k,i,i^\prime,j,j^\prime} \hat{P}_{i,i^\prime,j,j^\prime} \\
p_k &= \sum_{i,i^\prime,j,j^\prime}\nu_{k,i,i^\prime,j,j^\prime} p_{i,i^\prime,j,j^\prime} \\
\hat{Q}_l &= \sum_{i,j,\xi:p_{\mathrm{test}}(i,j,\xi)>0}  \beta_{l,i,j,\xi} \hat{Q}_{\xi,i,j} \\
q_l &= \sum_{i,j,\xi:p_{\mathrm{test}}(i,j,\xi)>0} \beta_{l,i,j,\xi} q_{\xi|i,j},
\end{aligned}
\end{equation}
where $\nu_{k,i,i^\prime,j,j^\prime}$ and $\beta_{l,i,j,\xi}$ are real coefficients. The relaxed SDP has the following form,
\begin{equation}\label{eq:relaxedprimalmdi}
\begin{aligned}
& \max_{\hat{G}} \mathrm{tr}(\hat{E}_{\mathrm{ph}} \hat{G}) \\
\mathrm{s.t.} &  \quad \hat{G} \succeq 0 \\
 & \mathrm{tr}(\hat{P}_k\otimes \hat{I}_{\Xi}\hat{G}) = p_k , \quad  k \in \{1,2,\cdots, m\} \\
 & \mathrm{tr}(\hat{Q}_l\hat{G}) = q_l , \quad  l \in \{1,2, \cdots, n\}\\
 & \mathrm{tr}(\hat{I}_{AB} \otimes \ket{\xi}\bra{\xi^\prime}_\Xi\hat{G}) = 0, \quad \xi \neq \xi^\prime.
\end{aligned}
\end{equation}
Compared with Eq.~\eqref{eq:primalrelax}, the only difference is the additional constraints for classical register. Considering the dual problem of Eq.~\eqref{eq:relaxedprimalmdi}, the target function is independent of the Lagrange multipliers for the classical register constraints, which means we can freely set these multipliers to be zero. Then the classical register constraints vanish and the dual problem of Eq.~\eqref{eq:relaxedprimalmdi} has the same form as Eq.~\eqref{eq:dual}. In the finite-size analysis of measurement-device-independent QKD protocols, one only need to replace the quantifies depending on $y$ and $b$ with those depending on $j$ and $\xi$, then follow the same procedure in Sec.~\ref{sec:finite}. For example, one should replace $\beta_{l,i,y,b}$, $p_{\mathrm{test}}(i,y,b)$ and $N_{\mathrm{test}}^{i,y,b}$ with $\beta_{l,i,j,\xi}$, $p_{\mathrm{test}}(i,j,\xi)$ and $N_{\mathrm{test}}^{i,j,\xi}$, respectively.

\section{Example: Phase-Matching QKD}\label{Sec:PMQKD}

To show how our method works in detail, we take the phase-matching MDI-QKD \cite{Ma2018phase,lin2018simple} as an example, which has been proved to surpass the linear key-rate bound \cite{takeoka2014fundamental,pirandola2017fundamental} both in asymptotic limit and finite-size case. A simplified phase-matching MDI-QKD protocol is proposed recently \cite{primaatmaja2019versatile} with asymptotic security analysis. Our result shows that such a protocol can also surpass the linear key-rate bound in finite-size case. In the following, a coherent state with a complex amplitude $\alpha$ is represented by a ket vector $\ket{\alpha}$ which is written by
\begin{equation}
\ket{\alpha} = e^{-\frac{|\alpha|^2}{2}}\sum_{n=0}^\infty \frac{\alpha^n}{\sqrt{n!}} \ket{n},
\end{equation}
in a Fock basis $\{\ket{n}\}_n$.

We first list the parameters of the protocol. The protocol dictates that Alice and Bob randomly prepare $X$ and $Y$ basis states of an optical pulse. The probability of choosing $X$ basis is $p_{\mathrm{basis},0}$. The mean photon numbers of $X$ and $Y$ basis states are $\mu_0$ and $\mu_1$, respectively. Alice generates a random bit $\alpha \in \{0,1\}$. The probability of $\alpha=0$ is $p_{\mathrm{aux,0}}$. The probability of determining a trash round is $p_{\mathrm{trash}}$. The estimations of actual experiment results prior to the experiment are $q_l^{\mathrm{nom}}$ $(l\in\{1,2,3,4\})$. We also use three security parameters, $\epsilon_{\mathrm{ph}}$, $\epsilon_{\mathrm{EC}}$, and $s_{\mathrm{PA}}$.
The protocol is defined as follows.

\textbf{Real Protocol}
\begin{enumerate}
	\item 
Alice(Bob) randomly prepares a quantum state $\ket{(-1)^{\kappa_{a(b)}} (i)^{\pi_{a(b)}} \sqrt{\mu_{\pi_{a(b)}}}}$ and sends it to an untrusted central node Charlie. Here $\kappa_{a(b)}\in\{0,1\}$ represents the raw key bit while $\pi_{a(b)} \in \{0,1\}$ represents the basis choice where $\pi_{a(b)} = 0$ corresponds to $X$ basis and $\pi_{a(b)} = 1$ corresponds to $Y$ basis. The probabilities of preparing $\ket{(-1)^{\kappa_{a(b)}} \sqrt{\mu_{\pi_{a(b)}}}}$
and $\ket{(-1)^{\kappa_{a(b)}} i \sqrt{\mu_{\pi_{a(b)}}}}$ are $p_{\mathrm{basis},0}/2$ and $(1-p_{\mathrm{basis},0})/2$, respectively. Alice generates a bit $\alpha \in \{0,1\}$ with a probability distribution $\{p_{\mathrm{aux,0}},1-p_{\mathrm{aux,0}}\}$. With probability $p_{\mathrm{trash}}$, Alice determines the round is a trash round.


\item The central node Charlie is supposed to perform a Bell measurement and announce the measurement result $\xi$. The result $\xi =0(1)$ corresponds to the detection of in(anti)-phase pulse pair and $\xi =2$ corresponds to the inconclusive result. If $\xi=1$, Bob flips his bit $\kappa_b$ as his new raw key bit.
 \item They repeat steps 1 and 2 for $N_{\mathrm{tot}}$ times.
	\item Alice announces the positions of the trash rounds, and Alice and Bob discard their raw key bits for those rounds.
				For each round that is not a trash round, Alice announces the value of $\alpha$. Both Alice and Bob announce their basis choices.
				Each round is labeled as \emph{signal} if they both choose $X$ basis ($\pi_a=\pi_b=0$), $\xi \neq 2$, and $\alpha=0$.
				Otherwise the round is \emph{test}.
				For signal rounds, Alice and Bob record the number of signal rounds $N_{\mathrm{sig}}$ and keep their raw key bits.
				For test rounds, they announce their raw key bits and record the following quantities: the number $N_{\mathrm{pass}}^X$ of successful detections when they both choose $X$ basis, the number $N_{\mathrm{pass}}^Y$ of successful detections when they both choose $Y$ basis, and  the number $N_{\mathrm{bit}}^X$ and $N_{\mathrm{bit}}^Y$ of rounds where bit error occurs on each basis. Then they compute the key length
				\begin{equation}
				K = N_{\mathrm{sig}}\left[1-h\left(\frac{M_{\mathrm{ph}}^U(N_{\mathrm{sig}},\{N_l\}_l,\epsilon_{\mathrm{ph}})}{N_{\mathrm{sig}}}\right)\right] -s_{\mathrm{PA}}.
				\end{equation}

	\item Alice and Bob perform the post-processing to generate their final keys. Here, direct reconciliation is used and hence Alice's 
				 sifted-key value is the reconciled sifted-key value.
							
\end{enumerate}
For convenience, we clarify how the protocol descriptions above satisfy the general prescription given in Sec.~\ref{sec:mdi}. In the protocol description, we introduced new labels $\kappa_{a(b)}$ and $\pi_{a(b)}$. The relations to the labels $i$ and $j$ are $i =(\kappa_a, \pi_a)=2\kappa_a + \pi_a +1$ and $j =(\kappa_b, \pi_b)=2\kappa_b + \pi_b +1$. Then the probabilities $\tau_A^i$ and $\tau_B^j$ satisfy $\tau_A^1=\tau_A^2=\tau_B^1=\tau_B^2 = p_{\mathrm{basis},0}/2$, and $\tau_A^3=\tau_A^4=\tau_b^3=\tau_b^4 =(1-p_{\mathrm{basis},0})/2$. We also introduces new random variables $N_{\mathrm{bit}}^X$, $N_{\mathrm{bit}}^Y$, $N_{\mathrm{pass}}^X$ and $N_{\mathrm{pass}}^Y$. The relations to $N_l$ will be given later.
The announcement is represented by $\tilde{\gamma} = f^\xi_{\mathrm{ann}}(i,j,\alpha)\coloneqq ((i\;\mathrm{mod}\;2),(j\;\mathrm{mod}\;2),\alpha)$. The round is signal when $\tilde{\gamma} \in W^\xi \coloneqq \{(1,1,0)\}$ for $\xi \in \{ 0,1\}$. Note that $W^2 =\emptyset$. We then have $\mathcal{W}^\xi=\{ \Omega_{(1,1,0)}\}$ $(\xi \in \{ 0,1\})$ with $\Omega_{(1,1,0)}=\{(i,j,0) | i\in \{1,3\}, j\in\{1,3\}\}$ and $\mathcal{W}^2=\emptyset$. The function $f^\xi_{\mathrm{rec}}(i,j,\alpha)$ is defined for $\xi \in \{ 0,1\}$ and $(i,j,\alpha)\in \Omega_{(1,1,0)}$,
				\begin{equation}
				 f^\xi_{\mathrm{rec}}(i,j,\alpha) =
					\begin{cases}
					 0, & i=1, j\in\{1,3\}, \alpha=0\\
					 1, & i=3, j\in\{1,3\}, \alpha=0\\ 
					\end{cases}.
				\end{equation}

To define the phase error operator, we consider the corresponding entanglement-based protocol. The initial pure states prepared by Alice and Bob, which was given in Eq.~\eqref{eq:mdidensity} as a total density operator for ${\cal H}_{A_0} \otimes {\cal H}_{B_0}  \otimes {\cal H}_{A_1} \otimes {\cal H}_{B_1}$, are represented by
\begin{equation}
\begin{aligned}
&\frac{1}{\sqrt{2}}\sum_{\kappa_{a}=0}^1\sum_{\pi_{a}=0}^1 \sqrt{p_{\mathrm{basis},\pi_a}} \ket{\kappa_a}_{A_0}\ket{\pi_a}_{A_1} \ket{(-1)^{\kappa_a}(i)^{\pi_a}\sqrt{\mu_{\pi_a}}}_{A^\prime} \\
&\frac{1}{\sqrt{2}}\sum_{\kappa_{b}=0}^1\sum_{\pi_{b}=0}^1 \sqrt{p_{\mathrm{basis},\pi_b}} \ket{\kappa_b}_{B_0}\ket{\pi_b}_{B_1} \ket{(-1)^{\kappa_b}(i)^{\pi_b}\sqrt{\mu_{\pi_b}}}_{B^\prime},
\end{aligned}
\end{equation}
where ${\cal H}_A= {\cal H}_{A_0} \otimes {\cal H}_{A_1}$ and ${\cal H}_B= {\cal H}_{B_0} \otimes {\cal H}_{B_1}$, such that ${\cal H}_{A_0}$, ${\cal H}_{A_1}$, ${\cal H}_{B_0}$, and ${\cal H}_{B_1}$ hold Alice's key bit, Alice's basis, Bob's key bit, and Bob's basis, respectively. The central node Charlie performs a Bell measurement on $\mathcal{H}_{A^\prime} \otimes \mathcal{H}_{B^\prime}$ and announce the measurement result $\xi$. The result $\xi =0(1)$ corresponds to the detection of in(anti)-phase pulse pair and $\xi =2$ corresponds to the inconclusive result. The POVM element $E^{\mathrm{obs},\xi}_{\mathrm{bit},z,\gamma}$ $(\gamma\in {\cal W})$ is given by Eq.~\eqref{eq:povmmdi}, which leads to 
\begin{equation}
\begin{aligned}
E^{\mathrm{obs},\xi}_{\mathrm{bit},z,\gamma} =  p_{\mathrm{aux},0} \ket{z}\bra{z}_{A_0}\otimes \hat{I}_{B_0} \otimes \ket{00}\bra{00}_{A_1B_1},\quad \xi\in\{0,1\}.
\end{aligned}
\end{equation}

Based on the asymptotic analysis of the protocol \cite{primaatmaja2019versatile} we define the following CP instrument $\{\zeta_\gamma^\xi\}_\gamma$. Bob performs $\hat{X}$ operation on ${\cal H}_{B_0}$ when $\xi=1$. Alice and Bob perform a $\{\ket{0}\bra{0},\ket{1}\bra{1}\}$ measurement on ${\cal H}_{A_1}$ and ${\cal H}_{B_1}$ and announce their results $\pi_a$ and $\pi_b$. If $\pi_a=\pi_b=0$ and $\xi \neq 2$, with probability $p_{\mathrm{aux,0}}$, Alice and Bob perform operation $\hat{S}^\dag_{A_0}\hat{U}_ \mathrm{CY}$ on ${\cal H}_{A_0} \otimes {\cal H}_{B_0}$, where $\hat{U}_\mathrm{CY}$ is the controlled-$\hat{Y}$ operation given by $\ket{0}\bra{0}_{A_0}\otimes \hat{I}_{B_0} + \ket{1}\bra{1}_{A_0} \otimes \hat{Y}_{B_0}$. Otherwise, they project ${\cal H}_{A_0}$ to $\ket{0}$. Finally they rename ${\cal H}_{A_0}$ as ${\cal H}_{Q}$. Here $\hat{Y} = -i\ket{0}\bra{1}+i\ket{1}\bra{0}$ is the Pauli operator and $\hat{S} = \ket{0}\bra{0}+i\ket{1}\bra{1}$ is the phase operator. One can easily check the CP instrument followed by a projection $\ket{z}\bra{z}$ $(z\in\{0,1\})$ on ${\cal H}_{Q}$ equals to the POVM element $E^{\mathrm{obs},\xi}_{\mathrm{bit},z,\gamma}$ $(\gamma\in {\cal W})$. Therefore, the outcomes of the measurement $\{\ket{0}\bra{0},\ket{1}\bra{1}\}$ on ${\cal H}_{Q}$ reproduce the joint probability distribution of the reconciled sifted-key value. The phase-control protocol can also be constructed according to the CP instrument.

The phase error operator $\hat{E}_{\mathrm{ph}}^{\mathrm{obs},\xi}$ is defined as $\sum_{\gamma}(\zeta^\xi_{\gamma})^\dagger(\ket{-}\bra{-})$ based on the CP instrument, which satisfies
\begin{equation}
\begin{aligned}
\hat{E}_{\mathrm{ph}}^{\mathrm{obs},0} & = p_{\mathrm{aux},0}
 \hat{U}_\mathrm{CY}^\dag \hat{S}_{A_0} \ket{-}\bra{-}_{A_0} \hat{S}^\dag_{A_0} \hat{U}_\mathrm{CY} \otimes \ket{00}\bra{00}_{A_1B_1}\\
   \hat{E}_{\mathrm{ph}}^{\mathrm{obs},1} & =  p_{\mathrm{aux},0}
\hat{X}_{B_0}
 \hat{U}_\mathrm{CY}^\dag \hat{S}_{A_0} \ket{-}\bra{-}_{A_0} \hat{S}^\dag_{A_0} \hat{U}_\mathrm{CY} \hat{X}_{B_0} \otimes \ket{00}\bra{00}_{A_1B_1}\\
		\hat{E}_{\mathrm{ph}}^{\mathrm{obs},2} & = 0,
\end{aligned}
\end{equation}
where we omit the identity operators. After some derivations, we can get the renormalized phase error operator,
\begin{widetext}
\begin{equation}\label{eq:phaseoperator}
\begin{aligned}
\hat{E}_{\mathrm{ph}} & =  \sum_{\xi=0}^2\mathcal{T}^{-1}_{\mathrm{MDI}}(\hat{E}_{\mathrm{ph}}^{\mathrm{obs},\xi}) \otimes \ket{\xi}\bra{\xi}_\Xi \\
& = \frac{p_{\mathrm{aux},0}p_{\mathrm{basis},0}^2}{4} \sum_{\xi=0}^1 \hat{X}_{B_0}\left(\frac{\hat{I}_{A_0}
\otimes \hat{I}_{B_0} - \hat{Y}_{A_0}
\otimes \hat{Y}_{B_0}}{2}\right) \hat{X}_{B_0} \otimes \ket{00}\bra{00}_{A_1B_1}\otimes \ket{\xi}\bra{\xi}_\Xi. \\
\end{aligned}
\end{equation}
\end{widetext}

We define operators $\hat{Q}_l$ $(l=1,2,3,4)$ and $\hat{P}_k$ by heuristically choosing the coefficients in Eq.~\eqref{eq:relaxationMDI} for the SDP problem,
\begin{widetext}
\begin{equation}
\begin{aligned}
\hat{Q}_1  = \hat{E}_{\mathrm{bit}}^X & =\frac{1}{4} \sum_{\xi=0}^1 \hat{X}_{B_0}^\xi \left( \frac{\hat{I}_{A_0}
\otimes \hat{I}_{B_0} - \hat{Z}_{A_0}
\otimes \hat{Z}_{B_0}}{2}\right) \hat{X}_{B_0}^\xi \otimes \ket{00}\bra{00}_{A_1B_1} \otimes \ket{\xi}\bra{\xi}_\Xi\\ 
&= \frac{1}{4}\sum_{\kappa_a,\pi_a,\kappa_b,\pi_b,\xi}\delta_{\pi_a ,0}\delta_{\pi_b,0}
 (1-\delta_{\kappa_a, (\kappa_b\oplus\xi)})(1-\delta_{\xi,2})\hat{Q}_{(\kappa_a,\pi_a),(\kappa_b,\pi_b),\xi} \\
\hat{Q}_2 = \hat{E}_{\mathrm{bit}}^Y   & = \frac{1}{4} \sum_{\xi=0}^1 \hat{X}^\xi_{B_0} \left(\frac{ \hat{I}_{A_0}
\otimes \hat{I}_{B_0} - \hat{Z}_{A_0}
\otimes \hat{Z}_{B_0}}{2}\right) \hat{X}^\xi_{B_0} \otimes \ket{11}\bra{11}_{A_1B_1} \otimes \ket{\xi}\bra{\xi}_\Xi\\
& =  \frac{1}{4}\sum_{\kappa_a,\pi_a,\kappa_b,\pi_b,\xi}\delta_{\pi_a ,1}\delta_{\pi_b,1} 
 (1-\delta_{\kappa_a, (\kappa_b\oplus\xi)})(1-\delta_{\xi,2})\hat{Q}_{(\kappa_a,\pi_a),(\kappa_b,\pi_b),\xi} \\
\hat{Q}_3 = \hat{P}_{\mathrm{pass}}^X & = \frac{1}{4} \sum_{\xi=0}^1  \hat{I}_{A_0}
\otimes \hat{I}_{B_0} \otimes \ket{00}\bra{00}_{A_1B_1} \otimes \ket{\xi}\bra{\xi}_\Xi\\
&= \frac{1}{4}\sum_{\kappa_a,\pi_a,\kappa_b,\pi_b,\xi}\delta_{\pi_a ,0}\delta_{\pi_b,0}  
 (1-\delta_{\xi,2})
 \hat{Q}_{(\kappa_a,\pi_a),(\kappa_b,\pi_b),\xi} \\
\hat{Q}_4 = \hat{P}_{\mathrm{pass}}^Y & = \frac{1}{4} \sum_{\xi=0}^1  \hat{I}_{A_0}
\otimes \hat{I}_{B_0} \otimes \ket{11}\bra{11}_{A_1B_1} \otimes \ket{\xi}\bra{\xi}_\Xi\\ \\
& =\frac{1}{4}\sum_{\kappa_a,\pi_a,\kappa_b,\pi_b,\xi}\delta_{\pi_a ,1}\delta_{\pi_b,1}
(1-\delta_{\xi,2})
 \hat{Q}_{(\kappa_a,\pi_a),(\kappa_b,\pi_b),\xi} \\ \\
\hat{P}_k & =  \sum_{\substack{\kappa_a,\kappa_b,\pi\\\kappa_a^\prime,\kappa_b^\prime,\pi^\prime}}  \hat{P}_{(\kappa_a,\pi),(\kappa_b,\pi), (\kappa_a^\prime,\pi^\prime),(\kappa_b^\prime,\pi^\prime)}, \quad k = 1,2,3,\cdots 64
\end{aligned}
\end{equation}
\end{widetext}
where $\hat{X}$ and $\hat{Z}$ are Pauli operators. We only consider the $64$ inner products of the same basis states, $k = 1+\kappa_a+2\pi+2^2\kappa_b+2^3(\kappa_a^\prime+2\pi^\prime+2^2\kappa_b^\prime)$.

Then one can substitute the operators into the primal SDP to calculate the asymptotic key rate. The explicit form of the SDP is given in Appendix~\ref{app:SDP}.
The dual problem is given by
\begin{equation}\label{eq:PMQKDdual}
\begin{aligned}
& \min_{\vec{\Lambda}} - \vec{\Lambda} \cdot \vec{C} \\
\mathrm{s.t.} \quad & \hat{E}_{\mathrm{ph}} + \eta_1 \hat{E}_{\mathrm{bit}}^X + \eta_2 \hat{E}_{\mathrm{bit}}^Y + \eta_3 \hat{P}_{\mathrm{pass}}^X + \eta_4 \hat{P}_{\mathrm{pass}}^Y  \\ &+ \sum_{k=1}^{64} \lambda_k \hat{P}_k \otimes \hat{I}_\Xi \preceq 0.
\end{aligned}
\end{equation}      
The real vector $\vec{C}$ is given by $\vec{C}=(p_1, p_2, \cdots, p_{64}, q_1^{\mathrm{nom}},q_2^{\mathrm{nom}},q_3^{\mathrm{nom}},q_4^{\mathrm{nom}})$, where $q_1^{\mathrm{nom}} =p_{\mathrm{pass}}^{X,\mathrm{nom}}e_{\mathrm{bit}}^{X,\mathrm{nom}}$, $q_2^{\mathrm{nom}} =p_{\mathrm{pass}}^{Y,\mathrm{nom}} e_{\mathrm{bit}}^{Y,\mathrm{nom}}$, $q_3^{\mathrm{nom}}= p_{\mathrm{pass}}^{X,\mathrm{nom}}$ and $q_4^{\mathrm{nom}} =p_{\mathrm{pass}}^{Y,\mathrm{nom}}$ are our estimations of the experimental results by assuming a channel model. In Appendix~\ref{app:simulation}, we give an example of estimations. By solving the dual problem, we have the following operator inequality
\begin{equation}\label{eq:operatorbound}
\begin{aligned}
& \hat{E}_{\mathrm{ph}} + \eta_1^* \hat{E}_{\mathrm{bit}}^X + \eta_2^* \hat{E}_{\mathrm{bit}}^Y + \eta_3^* \hat{P}_{\mathrm{pass}}^X + \eta_4^*  \hat{P}_{\mathrm{pass}}^Y   \\ &+\sum_{k=1}^{64} \lambda_k^* \hat{P}_k \otimes \hat{I}_\Xi \preceq 0,
\end{aligned}
\end{equation}
and the inequality for conditional expectations,
\begin{equation}\label{eq:expecineq}
\begin{aligned}
& \frac{1}{1-p_{\mathrm{trash}}}E(\chi_{\mathrm{ph}}^{(u)}|F^{u-1})+ \frac{1}{1-p_{\mathrm{trash}}}\sum_{l=1}^4 \eta_l^* E(\chi_{Q,l}^{(u)}|F^{u-1})\\ & +\frac{1}{p_{\mathrm{trash}}}E(\chi_{P}^{(u)}|F^{u-1})\leq 0.
\end{aligned}
\end{equation}

Properties specific to the current example can be used for reducing the computational cost of solving the SDP of Eq.~\eqref{eq:PMQKDdual}. We notice that $\hat{E}_{\mathrm{ph}}$ is proportional to $p_{\mathrm{aux},0}$ and $p_{\mathrm{basis,0}}^2$. If we consider another set of values $p^\prime_{\mathrm{aux},0}$ and $p^\prime_{\mathrm{basis,0}}$, then we can immediately have $\vec{\Lambda}^{*\prime} = p^\prime_{\mathrm{aux},0}(p^\prime_{\mathrm{basis,0}})^2\vec{\Lambda}^{*}/(p_{\mathrm{aux},0}p_{\mathrm{basis,0}}^2)$, where $\vec{\Lambda}^*=(\lambda_1^*, \lambda_2^*, \cdots, \lambda_m^*, \eta_1^*, \eta_2^*, \cdots, \eta_n^*)$. Therefore, a single run of the SDP is enough for different set of $\{p_{\mathrm{aux},0}, p_{\mathrm{basis,0}}\}$, which greatly saves the computation resource for the optimization of $p_{\mathrm{aux},0}$ and $p_{\mathrm{basis,0}}$. Moreover, we also notice that all the operators in Eq.~\eqref{eq:operatorbound} are block diagonal, i.e, they can be expressed as $\hat{O}_{AB\Xi} = \sum_{\xi=0}^2 \hat{O}_{AB}^\xi \otimes \ket{\xi}\bra{\xi}_\Xi$. If these operators satisfy $ \hat{O}_{AB}^0 = \hat{U} \hat{O}_{AB}^1 \hat{U}^\dag $ for some unitary operator $\hat{U}$, the constraint Eq.~\eqref{eq:operatorbound} can be further simplified into $\hat{E}_{\mathrm{ph}}^0 + \eta_1^* \hat{E}_{\mathrm{bit}}^{X,0} + \eta_2^* \hat{E}_{\mathrm{bit}}^{Y,0} + \eta_3^* \hat{P}_{\mathrm{pass}}^{X,0} + \eta_4^*  \hat{P}_{\mathrm{pass}}^{Y,0}   +\sum_{k=1}^{64} \lambda_k^* \hat{P}_k \preceq 0$.

Then we come to the finite-size analysis. Suppose the number of total rounds is $N_{\mathrm{tot}}$. 
We consider the random variables $\chi_{\mathrm{ph}}^{(u)}$, $\chi_{\mathrm{sig}}^{(u)}$, $\chi_{Q,l}^{(u)}$ $(l\in\{1,2,3,4\})$, and $\chi_{P}^{(u)}$ for $u$-th round $(u=1,2,\cdots,N_{\mathrm{tot}})$, which are defined in Table~\ref{tab:randomvariable}.
\begin{table*}[bt]
\begin{tabular}{c|c}
\hline
$\chi^{(u)}_{\mathrm{ph}}$ & Event of the $u$-th round \\
\hline
$1$ & signal round, phase error \\
$0$  & other cases\\
\hline
\hline
$\chi^{(u)}_{\mathrm{sig}}$ & Event of the $u$-th round \\
\hline
$1$ & signal round \\
$0$  & other cases\\
\hline
\hline
$\chi^{(u)}_{Q,1}$ & Event of the $u$-th round \\
\hline
$\frac{1}{p_{\mathrm{basis},0}^2p_{\mathrm{aux},1}}$  & test round, Alice and Bob both choose $X$ basis, successful detection, bit error \\
$0$  & other cases\\
\hline
\hline
$\chi^{(u)}_{Q,2}$ & Event of the $u$-th round \\
\hline
$\frac{1}{p_{\mathrm{basis},1}^2}$  & test round, Alice and Bob both choose $Y$ basis, successful detection, bit error \\
$0$  & other cases\\
\hline
\hline
$\chi^{(u)}_{Q,3}$ & Event of the $u$-th round \\
\hline
$\frac{1}{p_{\mathrm{basis},0}^2p_{\mathrm{aux},1}}$  & test round, Alice and Bob both choose $X$ basis,, successful detection\\
$0$  & other cases\\
\hline
\hline
$\chi^{(u)}_{Q,4}$ & Event of the $u$-th round \\
\hline
$\frac{1}{p_{\mathrm{basis},1}^2}$  & test round, Alice and Bob both choose $Y$ basis, successful detection\\
$0$  & other cases\\
\hline
$\chi^{(u)}_{P}$ & Event of the $u$-th round \\
\hline
$\omega_w$  & trash round, measurement result is $\omega_w$\\
$0$  & other cases\\
\hline
\end{tabular}
\caption{Correspondence between the values of random variables and events}
\label{tab:randomvariable}
\end{table*}
In finite-size analysis, we try to find $M_{\mathrm{ph}}^U(N_{\mathrm{sig}},\{N_l\}_l,\epsilon_{\mathrm{ph}})$. In the protocol, Alice and Bob record $N_{\mathrm{bit}}^X$, $N_{\mathrm{bit}}^Y$, $N_{\mathrm{pass}}^X$ and $N_{\mathrm{pass}}^Y$. Their relations to $N_l$ are given by
\begin{equation}
\begin{aligned}
N_{\mathrm{bit}}^X &=  p_{\mathrm{basis},0}^2p_{\mathrm{aux},1} N_1\\
N_{\mathrm{bit}}^Y &=  p_{\mathrm{basis},1}^2 N_2 \\
N_{\mathrm{pass}}^X &= p_{\mathrm{basis},0}^2p_{\mathrm{aux},1} N_3 \\
N_{\mathrm{pass}}^Y& = p_{\mathrm{basis},1}^2 N_4. \\
\end{aligned}
\end{equation}

We separately apply Kato's inequality to $\chi_{\mathrm{ph}}^{(u)}$, $\chi_{Q,l}^{(u)}$ $( l\in\{1,2,3,4\})$ and $\chi_{P}^{(u)}$ and have the following inequalities,
\begin{align}
\sum_{u=1}^{N_{\mathrm{tot}}}\chi^{(u)}_{\mathrm{ph}}
 & \leq  \sum_{u=1}^n E(\chi^{(u)}_{\mathrm{ph}}|F^{u-1}) 
  +\Delta^{1}_{\mathrm{ph}}(N_{\mathrm{tot}},N_{\mathrm{sig}},\epsilon_0) \label{eq:kato_chiph} \\  
\sum_{u=1}^{N_{\mathrm{tot}}} \eta_l^* \chi^{(u)}_{Q,l}
 & \leq  \sum_{u=1}^n  \eta_l^* E(\chi^{(u)}_{Q,l}|F^{u-1}) 
				+\Delta_{Q,l}^{\mathrm{sgn}(\eta_l^*)}(N_{\mathrm{tot}},N_l,\epsilon_l)  \label{eq:kato_chil}\\
\sum_{u=1}^{N_{\mathrm{tot}}}\chi^{(u)}_{P}				
 & \leq  \sum_{u=1}^n E(\chi^{(u)}_{P}|F^{u-1}) 
  + \Delta^{1}_{P}(N_{\mathrm{tot}}, \epsilon_5), \label{eq:kato_chip}
\end{align}
which hold with probabilities at least $1-\epsilon_0$, $1-\epsilon_l$ $(l\in\{1,2,3,4\})$ and $1-\epsilon_5$, respectively.  Here $\mathrm{sgn}(\cdot)=1(0)$ represents a positive(negative) value and $\Delta^{1(0)}$ is defined in Appendix~\ref{app:azuma}. The failure probabilities satisfy $\sum_{j=0}^5 \epsilon_j =\epsilon_{\mathrm{ph}}/2$. In Appendix~\ref{app:azuma}, we prove Eqs.~\eqref{eq:kato_chiph} to \eqref{eq:kato_chip}.  In Appendix~\ref{app:simulation}, we give an example of distributing these failure probabilities. Then $\Delta_1$ introduced in Eq.~\eqref{eq:conineq12} is given by
\begin{equation}
\begin{aligned}
& \Delta_1(N_{\mathrm{sig}},\{N_l\}_l,N_{\mathrm{tot}},\frac{\epsilon_\mathrm{ph}}{2})  \\ &= \frac{\Delta^1_{\mathrm{ph}}(N_{\mathrm{tot}},N_{\mathrm{sig}},\epsilon_0)  + \sum_{l=1}^4 \Delta_{Q,l}^{\mathrm{sgn}(\eta_l^*)}(N_{\mathrm{tot}},N_l,\epsilon_l)}{1-p_{\mathrm{trash}}} \\ 
& \phantom{=\;\;}+ \frac{\Delta^{1}_{P}(N_{\mathrm{tot}}, \epsilon_5)}{p_{\mathrm{trash}}}.
\end{aligned}
\end{equation}
We apply Bernstein's inequality to $\chi_{P}^{(u)}$ and have the following inequality,
\begin{equation}\label{bernsteinbound}
N_{\mathrm{tot}} \sum_{k=1}^{64} \lambda^*_k p_k -\Delta_2(N_{\mathrm{tot}},\frac{\epsilon_{\mathrm{ph}}}{2}) \leq \sum_{u=1}^{N_{\mathrm{tot}}}\chi_P^{(u)},
\end{equation}
which holds with probability at least $1-\epsilon_{\mathrm{ph}}/2$. Here $\Delta_2(N_{\mathrm{tot}},\epsilon_{\mathrm{ph}}/2) = \Delta_{\mathrm{Bern}}(N_{\mathrm{tot}},\epsilon_{\mathrm{ph}}/2)$. In Appendix~\ref{app:bernstein}, we define $\Delta_{\mathrm{Bern}}$ and prove Eq.~\eqref{bernsteinbound}. Combining Eqs.~\eqref{eq:expecineq} to \eqref{bernsteinbound}, we have
\begin{equation}
\label{eq:ChiPh_concentration_total}
\begin{aligned}
\sum_{u=1}^{N_{\mathrm{tot}}}\chi_{\mathrm{ph}}^{(u)} & \leq -\sum_{u=1}^{N_{\mathrm{tot}}}\sum_{l=1}^4 \eta_l^* \chi_{Q,l}^{(u)}- \frac{1-p_{\mathrm{trash}}}{p_{\mathrm{trash}}}N_{\mathrm{tot}} \sum_{k=1}^{64} \lambda_k^* p_k \\
& \phantom{=\;\;} + (1-p_{\mathrm{trash}}) \Delta_1 + \frac{1-p_{\mathrm{trash}}}{p_{\mathrm{trash}}}\Delta_2 \\
& = -\sum_{l=1}^4 \eta_l^* N_l- \frac{1-p_{\mathrm{trash}}}{p_{\mathrm{trash}}}N_{\mathrm{tot}} \sum_{k=1}^{64} \lambda_k^* p_k  \\ 
& \phantom{=\;\;}+ (1-p_{\mathrm{trash}})\Delta_1 + \frac{1-p_{\mathrm{trash}}}{p_{\mathrm{trash}}}\Delta_2,
\end{aligned}
\end{equation}
which holds with probability at least $1-\epsilon_{\mathrm{ph}}$.
Then $M_{\mathrm{ph}}^U(N_{\mathrm{sig}},\{N_l\}_l,\epsilon_{\mathrm{ph}})$ is given by the rhs of Eq.~\eqref{eq:ChiPh_concentration_total}.

\begin{figure}[hbt]
\centering
   \begin{minipage}[t]{0.48\textwidth}
   \centering
 \includegraphics[width=8cm]{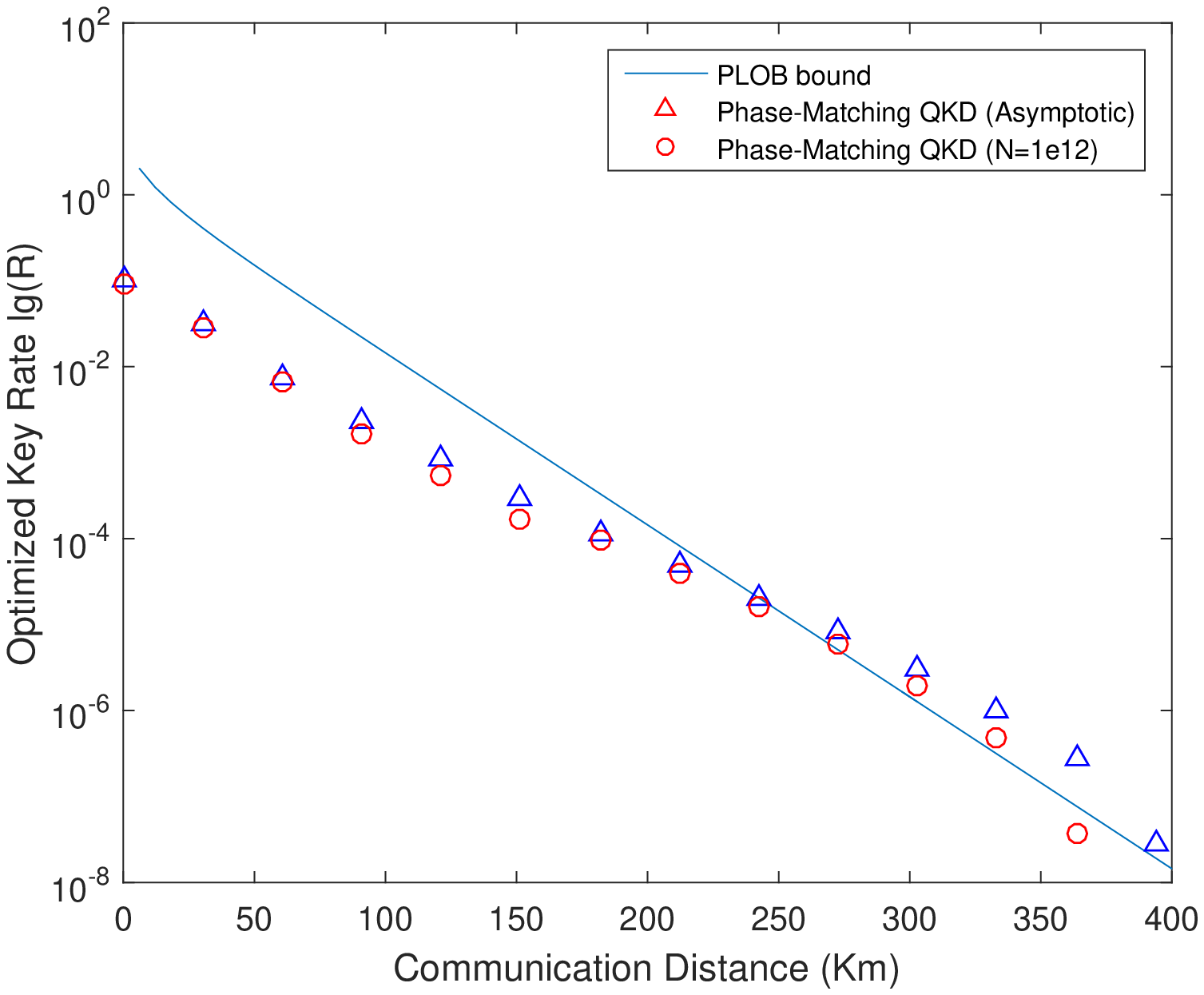}
   \end{minipage}
   \begin{minipage}[t]{0.48\textwidth}
   \centering
  \includegraphics[width=8cm]{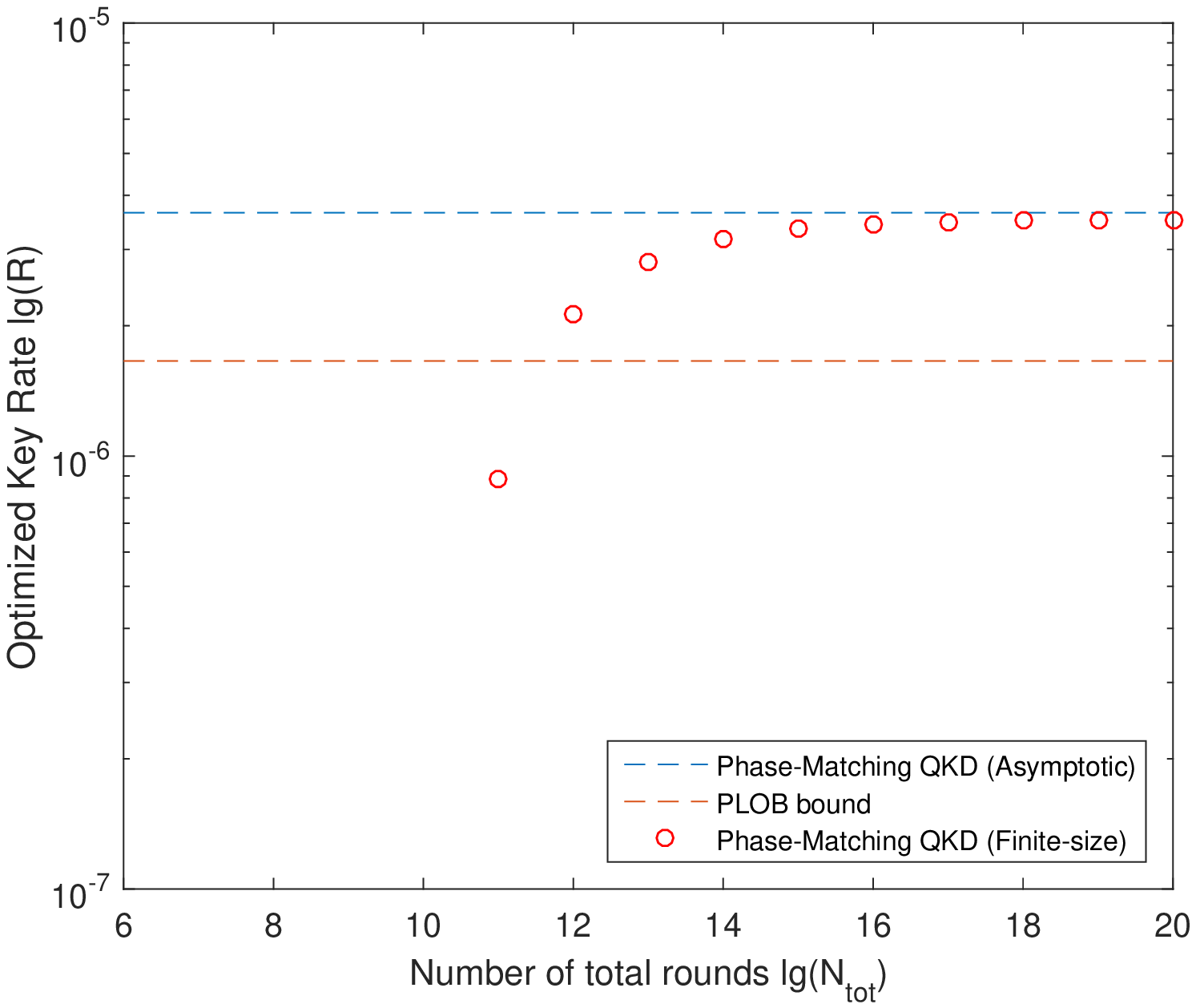}
   \end{minipage}

\caption{Simulation of the finite-size key rate versus the transmission distance (top) and the finite-size key rate versus the total rounds $N_{\mathrm{tot}}$ at $300$km (bottom). We assume the detector efficiency is $1$ and choose the dark count rate $p_d=10^{-8}$, the number of total rounds $N_{\mathrm{tot}}=10^{12}$, and security parameter $\epsilon_{\mathrm{ph}}=2^{-66}$. Here $\mu_1$ and $\mu_2$ are optimized.} 
\label{fig:simulation}
\end{figure}

The result enables us to simulate the finite-size key rate versus the transmission distance (Fig.~\ref{fig:simulation}). It turns out that such a protocol can also surpass the linear key-rate bound with $N_{\mathrm{tot}}=10^{12}$ rounds. The simulation formulas are given in Appendix~\ref{app:simulation}.

\section{Discussion and conclusion}

In conclusion, we give a numerical finite-size security analysis method for prepare-and-measure and measurement-device-independent QKD protocols. In the finite-size analysis, we introduce trash rounds where virtual measurements are performed to deal with the constraints independent of attacks from eavesdroppers.
We also discriminate nominal and observed experimental data, which serve as protocol parameters and random variables, respectively.
For a practical QKD implementation, one can estimate the measurement results and substitute them into the SDP to get a valid (might not be optimal) operator inequality. Then the upper bound of the phase error occurrence can be calculated by combining suitable concentration inequalities and hence the finite-size analysis is completed. Our method will be a useful tool to calculate the finite-size secret key rate under general attacks, especially for non-professional persons to calculate the secure key rate. 

There are some future directions to extend our results. One interesting direction is to take various device imperfections into consideration, such as MDI-QKD with partially characterized sources. Besides, we can also explore whether our method can be generalized to continuous variable QKD protocols especially the discrete-modulated coherent state protocols. 

\section*{Acknowledgments}
We thank T. Matsuura and I. Primaatmaja for enlightening discussions. This work was supported by CREST (Japan Science and Technology
Agency) JPMJCR1671; JSPS KAKENHI Grant Number JP18K13469.

\appendix
\section{Numerical method with tag}\label{app:tag}
Suppose Bob applies tagging method to reduce the dimension of his signal to finite. In the tagging method, the Gram matrix as well as the other operators should have a direct sum form. We denotes the label of a tag as $t$ and define a new phase error operator as
\begin{equation}
\tilde{E}_{\mathrm{ph}} = \sum_t \lambda^t \hat{\Omega^t} + \mu^t \hat{E}_{\mathrm{ph}}^t,
\end{equation}
where $\lambda^t$ and $\mu^t$ are heuristically chosen coefficients while $\hat{\Omega^t}$ and $\hat{E}_{\mathrm{ph}}^t$ are observables. Here $\mathrm{tr}(\hat{G} \hat{\Omega^t})=\omega^t$ and $\mathrm{tr}(\hat{G} \hat{E}_{\mathrm{ph}}^t)=e_{\mathrm{ph}}^t$ are the proportion and the phase error rate of the $t$ tag, respectively. These operators are given according to a specific protocol. Then we replace the phase error operator with the new one $\tilde{E}_{\mathrm{ph}}$ in the SDP problem. With a similar process, we can calculate an upper bound of $\sum_t \lambda^t \omega^t + \mu^t e_{\mathrm{ph}}^t$ in finite-size case. We denote this upper bound by $\tilde{e}_{\mathrm{ph}}^U$. Then Eve's information leakage is given by the following convex optimization problem,
\begin{equation}\label{eq:cvxwithtag}
\begin{aligned}
& \max_{\omega^t, e_{\mathrm{ph}}^t} \sum_t \omega^t h(e_{\mathrm{ph}}^t) \\
\mathrm{s.t.} \quad &  \sum_t \lambda^t \omega^t + \mu^t e_{\mathrm{ph}}^t \leq \tilde{e}_{\mathrm{ph}}^U.
\end{aligned}
\end{equation}

\section{Application of improved Azuma's inequality}\label{app:azuma}
The improved Azuma's inequality is given by the following theorem.
\begin{theorem}
\rm{(Kato's inequality \cite{kato2020concentration})} \textit{Let $\{X^{(u)}\}$ be a list of random variables, and $\{F^{u-1}\}$ be a filtration that identifies random variables $\{X^{(1)},X^{(2)},\cdots X^{(u-1)}\}$. Suppose $0\leq X^{(u)}\leq 1$ for any $u$, then for any $n\in \mathbb{N}$, $a \in \mathbb{R}$ and $b \geq0$, we have the relation}
\begin{equation}\label{eq:Kato1}
\begin{aligned}
&\mathrm{Pr}\left\{\sum_{u=1}^n \left(E(X^{(u)}|F^{u-1})-X^{(u)}\right) \right. \\
&\phantom{=\;\;} \geq \left. \left[b +a\left(\frac{2\sum_{u=1}^n X^{(u)}}{n}-1\right)\right]\sqrt{n} \right\} \\ 
&\leq \exp{\left(-\frac{2b^2-2a^2}{(1+\frac{4a}{3\sqrt{n}})^2}\right)}.
\end{aligned}
\end{equation}
\end{theorem}
In Kato's inequality, we choose the parameters $a$ and $b$ to make the inequality tight when the random variable $X=\sum_{u=1}^n X^{(u)}$ is
close to its estimation.
We denote this estimation as $X^{\mathrm{nom}}$.
For the optimal choice of $a$ and $b$, we solve the following optimization problem for Eq.~\eqref{eq:Kato1}
\begin{equation}\label{eq:katoopt}
\begin{aligned}
& \min \quad  b +a\left(\frac{2X^{\mathrm{nom}}}{n}-1\right) \\
\mathrm{s.t.} & \quad \exp{\left(-\frac{2b^2-2a^2}{(1+\frac{4a}{3\sqrt{n}})^2}\right)} = \epsilon\\
& \quad b \geq 0,
\end{aligned}
\end{equation}
where $\epsilon$ is the given failure probability satisfying $0<\epsilon\leq 1$. The optimization problem can be solved analytically \cite{lorenzo2019tight},
\begin{widetext}
\begin{equation}
\begin{aligned}
a_0^*(n,X^{\mathrm{nom}},\epsilon) & = \frac{27\sqrt{2}(n-2X^{\mathrm{nom}})\sqrt{-n^2\ln\epsilon (9X^{\mathrm{nom}}(n-X^{\mathrm{nom}})-2n\ln \epsilon)}+216\sqrt{n}X^{\mathrm{nom}}(n-X^{\mathrm{nom}})\ln \epsilon -48 n^{3/2} \ln^2\epsilon}{4(9n-8\ln\epsilon)(9X^{\mathrm{nom}}(n-X^{\mathrm{nom}})-2n\ln\epsilon)}  \\
b_0^*(n,X^{\mathrm{nom}},\epsilon) & = \frac{\sqrt{18a_0^*(n,X^{\mathrm{nom}},\epsilon)^2n- (16a_0^*(n,X^{\mathrm{nom}},\epsilon)^2 -24a_0^*(n,X^{\mathrm{nom}},\epsilon) \sqrt{n}+9n)\ln \epsilon}}{3\sqrt{2n}}.
\end{aligned}
\end{equation}
\end{widetext}

Then according to Eq.~\eqref{eq:Kato1}, we have the following inequality which holds with probability at least $1-\epsilon$,
\begin{equation}\label{eq:katobounds}
\begin{aligned}
  \sum_{u=1}^n E(X^{(u)}|F^{u-1})  -  \sum_{u=1}^n X^{(u)} \leq \Delta^0(n,X,X^{\mathrm{nom}},\epsilon) 
\end{aligned}
\end{equation}
where 
\begin{equation}\label{eq:katodelta}
\begin{aligned}
&\phantom{=\;\;} \Delta^0(n,X,X^{\mathrm{nom}},\epsilon) \\ 
&= \left[b_0^*(n,X^{\mathrm{nom}},\epsilon) + a_0^*(n,X^{\mathrm{nom}},\epsilon) \left(\frac{2X}{n}-1\right)\right]\sqrt{n}.
\end{aligned}
\end{equation}
We can obtain another inequality by replacing $X^{(u)} \rightarrow 1-X^{(u)}$ and $a \rightarrow -a$ in Eq.~\eqref{eq:Kato1} and $X^{\mathrm{nom}} \rightarrow n-X^{\mathrm{nom}}$ in Eq.~\eqref{eq:katoopt},
\begin{equation}
\begin{aligned}
\sum_{u=1}^n X^{(u)} - \sum_{u=1}^n E(X^{(u)}|F^{u-1}) \leq      \Delta^1(n,X,X^{\mathrm{nom}},\epsilon),
\end{aligned}
\end{equation}
where 
\begin{widetext}
\begin{equation}
\begin{aligned}
\Delta^1(n,X,X^{\mathrm{nom}},\epsilon) &= \left[b_1^*(n,X^{\mathrm{nom}},\epsilon) + a_1^*(n,X^{\mathrm{nom}},\epsilon) \left(\frac{2X}{n}-1\right)\right]\sqrt{n},\\
 a_1^*(n,X^{\mathrm{nom}},\epsilon) & = -a_0^*(n,n-X^{\mathrm{nom}},\epsilon)\\
 b_1^*(n,X^{\mathrm{nom}},\epsilon) & = \frac{\sqrt{18a_1^*(n,X^{\mathrm{nom}},\epsilon)^2n- (16a_1^*(n,X^{\mathrm{nom}},\epsilon)^2 +24a_1^*(n,X^{\mathrm{nom}},\epsilon) \sqrt{n}+9n)\ln \epsilon}}{3\sqrt{2n}}.
\end{aligned}
\end{equation}
\end{widetext}

In our case, we need to normalize $\chi^{(u)}_{\mathrm{ph}}$, $\chi^{(u)}_{Q,l}$ $(l=1,2,3,4)$ and $\chi_P^{(u)}$
to make their range $[0,1]$, which is the range of $X^{(u)}$ in Kato's inequality. We define normalized random variables $\theta^{(u)}_{\mathrm{ph}}$, $\theta^{(u)}_{Q,l}$ $(l\in\{1,2,3,4\})$ and $\theta^{(u)}_{P}$ and their summations $\Theta_{\mathrm{ph}} = \sum_{u=1}^{N_{\mathrm{tot}}} \theta^{(u)}_{\mathrm{ph}}$, $\Theta_{Q,l} = \sum_{u=1}^{N_{\mathrm{tot}}}\theta^{(u)}_{Q,l}$ $(l\in\{1,2,3,4\})$ and $\Theta_P =\sum_{u=1}^{N_{\mathrm{tot}}}\theta^{(u)}_{P}$. The normalizations are given by
\begin{equation}
\begin{aligned}
\theta^{(u)}_{\mathrm{ph}} & = \chi^{(u)}_{\mathrm{ph}} \\
\theta^{(u)}_{Q,1} &=  p_{\mathrm{basis},0}^2p_{\mathrm{aux},1} \chi^{(u)}_{Q,1}\\
\theta^{(u)}_{Q,2} &= p_{\mathrm{basis},1}^2 \chi^{(u)}_{Q,2} \\
\theta^{(u)}_{Q,3} &= p_{\mathrm{basis},0}^2p_{\mathrm{aux},1} \chi^{(u)}_{Q,3} \\
\theta^{(u)}_{Q,4} & = p_{\mathrm{basis},1}^2 \chi^{(u)}_{Q,4}\\
\theta^{(u)}_{P} & = \frac{\chi^{(u)}_{P}-\omega_{\mathrm{min}}}{\omega_{\mathrm{max}}-\omega_{\mathrm{min}}},
\end{aligned}
\end{equation}
where $\omega_{\mathrm{max}}$ and $\omega_{\mathrm{min}}$ are the maximum and minimum value of $\omega_w$, respectively. Then $\Theta_{Q,l}$ is proportional to $N_l$,
\begin{equation}
\begin{aligned}
\Theta_{Q,1} &=  p_{\mathrm{basis},0}^2p_{\mathrm{aux},1} N_1\\
\Theta_{Q,2} &= p_{\mathrm{basis},1}^2 N_2 \\
\Theta_{Q,3} &= p_{\mathrm{basis},0}^2p_{\mathrm{aux},1} N_3 \\
\Theta_{Q,4} & = p_{\mathrm{basis},1}^2 N_4.\\
\end{aligned}
\end{equation}

We apply Kato's inequality to each normalized random variable separately, where we choose the parameter $a$ and $b$ separately. For random variables $\theta_{Q,l}^{(u)}$ $(l\in\{1,2,3,4\})$,  
we make estimations for $\sum_{u=1}^{N_{\mathrm{tot}}}\theta^{(u)}_{Q,l}$ before experiment, denoted as $\Theta_{Q,l}^{\mathrm{nom}}$, to find optimal $a$ and $b$.  An example of the estimations are given in Appendix~\ref{app:simulation}. After the experiment, we can observe $\Theta_{Q,l}$. 
We apply Kato's inequality and have the following inequalities,
\begin{equation}
\begin{aligned}
 \sum_{u=1}^{N_{\mathrm{tot}}}\theta^{(u)}_{Q,l} - \sum_{u=1}^{N_{\mathrm{tot}}}E(\theta^{(u)}_{Q,l}|F^{u-1})& \leq  \Delta^1\left(N_{\mathrm{tot}}, \Theta_{Q,l},\Theta_{Q,l}^{\mathrm{nom}},\epsilon_l\right) \\
\sum_{u=1}^{N_{\mathrm{tot}}} E(\theta^{(u)}_{Q,l}|F^{u-1}) - \sum_{u=1}^{N_{\mathrm{tot}}}\theta^{(u)}_{Q,l} & \leq    \Delta^0\left(N_{\mathrm{tot}}, \Theta_{Q,l},\Theta_{Q,l}^{\mathrm{nom}}, \epsilon_l\right), 
\end{aligned}
\end{equation}
where each of them holds with probability at least $1-\epsilon_l$. Given the protocol parameters, 
by setting $\Delta_{Q,l}^{\mathrm{sgn}(\eta_l^*)}(N_{\mathrm{tot}},N_l,\epsilon_l)$ as
\begin{equation}
\begin{aligned}
 \Delta_{Q,1}^{\mathrm{sgn}(\eta_1^*)}(N_{\mathrm{tot}},N_1,\epsilon_1)
	& = \frac{|\eta_1^*|\Delta^{\mathrm{sgn}(\eta_1^*)}(N_{\mathrm{tot}},\Theta_{Q,1},\Theta_{Q,1}^{\mathrm{nom}},\epsilon_1)}{p_{\mathrm{basis},0}^2p_{\mathrm{aux},1}} \\
	 \Delta_{Q,2}^{\mathrm{sgn}(\eta_2^*)}(N_{\mathrm{tot}},N_2,\epsilon_2)
	& = \frac{|\eta_2^*|\Delta^{\mathrm{sgn}(\eta_2^*)}(N_{\mathrm{tot}},\Theta_{Q,2},\Theta_{Q,2}^{\mathrm{nom}},\epsilon_2)}{p_{\mathrm{basis},1}^2} \\
	 \Delta_{Q,3}^{\mathrm{sgn}(\eta_3^*)}(N_{\mathrm{tot}},N_3,\epsilon_3)
	& = \frac{|\eta_3^*|\Delta^{\mathrm{sgn}(\eta_3^*)}(N_{\mathrm{tot}},\Theta_{Q,3},\Theta_{Q,3}^{\mathrm{nom}},\epsilon_3)}{p_{\mathrm{basis},0}^2p_{\mathrm{aux},1}} \\
	 \Delta_{Q,4}^{\mathrm{sgn}(\eta_4^*)}(N_{\mathrm{tot}},N_4,\epsilon_4)
	& = \frac{|\eta_4^*|\Delta^{\mathrm{sgn}(\eta_4^*)}(N_{\mathrm{tot}},\Theta_{Q,4},\Theta_{Q,4}^{\mathrm{nom}},\epsilon_4)}{p_{\mathrm{basis},1}^2}, \\
\end{aligned}
\end{equation}
we have
\begin{equation}
  \sum_{u=1}^{N_{\mathrm{tot}}}\eta_l^* \chi^{(u)}_{Q,l}
 \leq  \sum_{u=1}^{N_{\mathrm{tot}}} \eta_l^* E(\chi^{(u)}_{Q,l}|F^{u-1}) 
				+\Delta_{Q,l}^{\mathrm{sgn}(\eta_l^*)}(N_{\mathrm{tot}},N_l,\epsilon_l),
\end{equation}
which holds with probability at least $1-\epsilon_l$. 

The random variable $\Theta_{\mathrm{ph}}$ cannot be obtained in experiment. We use an upper bound $\Theta_{\mathrm{ph}} \leq N_{\mathrm{sig}}$ derived from the relation $\chi_{\mathrm{ph}}^{(u)}\leq \chi_{\mathrm{sig}}^{(u)}$.
Before experiment, we set the expectation of $\Theta_{\mathrm{ph}}$ to be $N_{\mathrm{sig}}^{\mathrm{nom}}$ and calculate $a^{*}_{1}(N_{\mathrm{tot}},N_{\mathrm{sig}}^{\mathrm{nom}},\epsilon_0)$ and $b^{*}_{1}(N_{\mathrm{tot}},N_{\mathrm{sig}}^{\mathrm{nom}},\epsilon_0)$.
We consider two different cases. When $a^{*}_{1}(N_{\mathrm{tot}},N_{\mathrm{sig}}^{\mathrm{nom}},\epsilon_0)>0$, we find
\begin{equation}
\begin{split}
 & \Delta^{1}(N_{\mathrm{tot}},\Theta_{\mathrm{ph}},N_{\mathrm{sig}}^{\mathrm{nom}},\epsilon_0) \\
 =& \left[b^*_1(N_{\mathrm{tot}},N_{\mathrm{sig}}^{\mathrm{nom}},\epsilon_0)+a^*_1(N_{\mathrm{tot}},N_{\mathrm{sig}}^{\mathrm{nom}},\epsilon_0)\left(\frac{2\Theta_{\mathrm{ph}}}{N_{\mathrm{tot}}}-1\right)\right]\sqrt{n}\\
 \leq& \left[ b^*_1(N_{\mathrm{tot}},N_{\mathrm{sig}}^{\mathrm{nom}},\epsilon_0)+a^*_1(N_{\mathrm{tot}},N_{\mathrm{sig}}^{\mathrm{nom}},\epsilon_0)\left(\frac{2N_{\mathrm{sig}}}{N_{\mathrm{tot}}}-1\right)\right]\sqrt{n}\\
 =&\Delta^{1}(N_{\mathrm{sig}},N_{\mathrm{sig}}^{\mathrm{nom}},\epsilon_0).
\end{split}
\end{equation}
Note that it mostly holds when $N_{\mathrm{tot}}$ is much larger than $|\log(\epsilon_{\mathrm{ph}})|$, and $N_{\mathrm{sig}}/N_{\mathrm{tot}}$ is less than $1/2$.
In this case, we set $\Delta^{1}_{\mathrm{ph}}(N_{\mathrm{tot}},N_{\mathrm{sig}},\epsilon_0) = \Delta^{1}(N_{\mathrm{tot}},N_{\mathrm{sig}},N_{\mathrm{sig}}^{\mathrm{nom}},\epsilon_0)$.
When $a^{*}_{1}(N_{\mathrm{tot}},N_{\mathrm{sig}}^{\mathrm{nom}},\epsilon_0)\leq 0$,
we newly choose $a$ and $b$ in Kato's inequality as $a=0$ and $b=\sqrt{-\log(\epsilon_0)/2}$.
It guarantees a slightly better bound than Azuma's inequality.
We set $\Delta^{1}_{\mathrm{ph}}(N_{\mathrm{tot}},N_{\mathrm{sig}},\epsilon_0) = \sqrt{-n\log(\epsilon_0)/2}$ based on this refined Azuma's inequality.
In both cases, we find the inequality 
\begin{equation}
  \sum_{u=1}^{N_{\mathrm{tot}}}\chi^{(u)}_{\mathrm{ph}}
 \leq  \sum_{u=1}^{N_{\mathrm{tot}}} E(\chi^{(u)}_{\mathrm{ph}}|F^{u-1}) 
  +\Delta^{1}_{\mathrm{ph}}(N_{\mathrm{tot}},N_{\mathrm{sig}},\epsilon_0),
\end{equation}
which holds with the probability at least $1-\epsilon_0$. Then we proved Eq.~\eqref{eq:kato_chiph}

As for $\Theta_{P}$, we use bounds of $\sum_{u=1}^{N_{\mathrm{tot}}}\chi_P^{(u)}$ deduced from Bernstein's inequality introduced in Appendix \ref{app:bernstein}
\begin{equation}\label{eq:katoconstraint}
\begin{aligned}
\sum_{u=1}^{N_{\mathrm{tot}}}\chi_P^{(u)} \leq N_{\mathrm{tot}} \sum_{k=1}^{64} \lambda^*_k p_k +\Delta_{\mathrm{Bern}}(N_{\mathrm{tot}},\epsilon_6) \\
\sum_{u=1}^{N_{\mathrm{tot}}}\chi_P^{(u)} \geq N_{\mathrm{tot}} \sum_{k=1}^{64} \lambda^*_k p_k -\Delta_{\mathrm{Bern}}(N_{\mathrm{tot}},\epsilon_6),
\end{aligned}
\end{equation}
where each of them holds with probability at least $1-\epsilon_6$.
Then we have the following bounds for $\Theta_P$,
\begin{equation}
\begin{aligned}
 \Theta_P &\leq \frac{N_{\mathrm{tot}}\sum_{k=1}^{64} \lambda^*_k p_k+\Delta_{\mathrm{Bern}}(N_{\mathrm{tot}},\epsilon_6) - N_{\mathrm{tot}}\omega_{\mathrm{min}}}{\omega_{\mathrm{max}}-\omega_{\mathrm{min}}} \\ 
 &=: \Theta_P^U  \\
 \Theta_P &\geq \frac{N_{\mathrm{tot}}\sum_{k=1}^{64} \lambda^*_k p_k-\Delta_{\mathrm{Bern}}(N_{\mathrm{tot}},\epsilon_6) -N_{\mathrm{tot}}\omega_{\mathrm{min}}}{\omega_{\mathrm{max}}-\omega_{\mathrm{min}}} \\ 
 &=: \Theta_P^L, 
\end{aligned}
\end{equation}
where each of them holds with probability at least $1-\epsilon_6$.
We make an estimation of $\Theta_P$, denoted as $\Theta_{P}^{\mathrm{nom}}=N_{\mathrm{tot}}(\sum_{k=1}^{64} \lambda^*_k p_k -\omega_{\mathrm{min}})/(\omega_{\mathrm{max}}-\omega_{\mathrm{min}})$, and calculate $a^{*}_{1}(N_{\mathrm{tot}},\Theta_{P}^{\mathrm{nom}},\epsilon_7)$ and $b^{*}_{1}(N_{\mathrm{tot}},\Theta_{P}^{\mathrm{nom}},\epsilon_7)$, where $\epsilon_6+\epsilon_7=\epsilon_5$. Similar to $\theta^{(u)}_{\mathrm{ph}}$, we will consider two different cases.
When $a_{1}^*(N_{\mathrm{tot}},\Theta_{P}^{\mathrm{nom}},\epsilon_7) >0$, we find 
\begin{equation}
\begin{split}
 & \Delta^{1}(N_{\mathrm{tot}},\Theta_{P},\Theta_{P}^{\mathrm{nom}},\epsilon_7) \\
	=& \left[b^*_1(N_{\mathrm{tot}},\Theta_{P}^{\mathrm{nom}},\epsilon_7)+a^*_1(N_{\mathrm{tot}},\Theta_{P}^{\mathrm{nom}},\epsilon_7)\left(\frac{2\Theta_{P}}{N_{\mathrm{tot}}}-1\right)\right]\sqrt{n} \\
	\leq& \left[b^*_1(N_{\mathrm{tot}},\Theta_{P}^{\mathrm{nom}},\epsilon_7)+a^*_1(N_{\mathrm{tot}},\Theta_{P}^{\mathrm{nom}},\epsilon_7)
 \left( 
 \frac{2\Theta_{P}^U}{N_{\mathrm{tot}}}-1\right)\right] \sqrt{n}\\
 =&\Delta^{1}(N_{\mathrm{tot}},\Theta_{P}^U,\Theta_{P}^{\mathrm{nom}},\epsilon_7).
\end{split}
\end{equation}
In this case, we set $\Delta^{1}_{P}(N_{\mathrm{tot}},\epsilon_7)
= (\omega_{\mathrm{max}}-\omega_{\mathrm{min}})\Delta^{1}(N_{\mathrm{tot}},\Theta_{P}^U,\Theta_{P}^{\mathrm{nom}},\epsilon_7)$.
When $a_{1}^*(N_{\mathrm{tot}},\Theta_{P}^{\mathrm{nom}},\epsilon_7)\leq 0$, we find 
\begin{equation}
\begin{split}
 & \Delta^{1}(N_{\mathrm{tot}},\Theta_{P},\Theta_{P}^{\mathrm{nom}},\epsilon_7) \\
	=& \left[b^*_1(N_{\mathrm{tot}},\Theta_{P}^{\mathrm{nom}},\epsilon_7)+a^*_1(N_{\mathrm{tot}},\Theta_{P}^{\mathrm{nom}},\epsilon_7)\left(\frac{2\Theta_{P}}{N_{\mathrm{tot}}}-1\right)\right]\sqrt{n} \\
	\leq& \left[b^*_1(N_{\mathrm{tot}},\Theta_{P}^{\mathrm{nom}},\epsilon_7)+a^*_1(N_{\mathrm{tot}},\Theta_{P}^{\mathrm{nom}},\epsilon_7)
 \left( 
 \frac{2\Theta_{P}^L}{N_{\mathrm{tot}}}-1\right)\right] \sqrt{n}\\
 =&\Delta^{1}(N_{\mathrm{tot}},\Theta_{P}^L,\Theta_{P}^{\mathrm{nom}},\epsilon_7).
\end{split}
\end{equation}
We set $\Delta^{1}_{P}(N_{\mathrm{tot}},\epsilon_7) = (\omega_{\mathrm{max}}-\omega_{\mathrm{min}}) \Delta^{1}(\epsilon_7,\Theta_{P}^L,\Theta_{P}^{\mathrm{nom}})$. In summary, $\Delta^{1}_{P}(N_{\mathrm{tot}},\epsilon_7)$ is defined as
\begin{equation}
\begin{aligned}
\Delta^{1}_{P}(N_{\mathrm{tot}},\epsilon_7) =\begin{cases}
					& (\omega_{\mathrm{max}}-\omega_{\mathrm{min}})\Delta^{1}(N_{\mathrm{tot}},\Theta_{P}^U,\Theta_{P}^{\mathrm{nom}},\epsilon_7), \\ & \phantom{=\;\;}\phantom{=\;\;}a_{1}^*(N_{\mathrm{tot}},\Theta_{P}^{\mathrm{nom}},\epsilon_7)> 0\\
					& (\omega_{\mathrm{max}}-\omega_{\mathrm{min}})\Delta^{1}(N_{\mathrm{tot}},\Theta_{P}^L,\Theta_{P}^{\mathrm{nom}},\epsilon_7), \\ & \phantom{=\;\;}\phantom{=\;\;}a_{1}^*(N_{\mathrm{tot}},\Theta_{P}^{\mathrm{nom}},\epsilon_7)\leq 0\\ 
					\end{cases}.
\end{aligned}
\end{equation}
We find the inequality 
\begin{equation}
  \sum_{u=1}^{N_{\mathrm{tot}}}\chi^{(u)}_{P}
 \leq  \sum_{u=1}^{N_{\mathrm{tot}}} E(\chi^{(u)}_{P}|F^{u-1}) 
  + \Delta^{1}_{P}(N_{\mathrm{tot}}, \epsilon_5),
\end{equation}
which holds with the probability at least $1-\epsilon_5$. Then we proved Eq.~\eqref{eq:kato_chip}.



\section{Application of Bernstein's inequality}\label{app:bernstein}
The Bernstein's inequality is given by the following theorem.
\begin{theorem}
\rm{(Bernstein's inequality \cite{bernstein1924modification})} \textit{Let $X^{(1)}$, $X^{(2)}$, $\cdots$, $X^{(n)}$ be independent zero-mean variables, i.e., $E(X^{(u)})=0$ $(u=1,2,\cdots,n)$. Suppose $-1 \leq X^{(u)} \leq 1$ $(u=1,2,\cdots,n)$, Then for all $\Delta> 0$,}
\begin{equation}\label{eq:bernstein}
\begin{aligned}
\mathrm{Pr}\left\{\frac{1}{n}\sum_{u=1}^n X^{(u)} \geq \Delta \right\} & \leq \exp\left(\frac{-n \Delta^2}{\frac{2}{n}\sum_{u=1}^n E((X^{(u)})^2) +\frac{2}{3}\Delta}\right) \\
\mathrm{Pr}\left\{\frac{1}{n}\sum_{u=1}^n X^{(u)} \leq -\Delta \right\} & \leq \exp\left(\frac{-n \Delta^2}{\frac{2}{n}\sum_{u=1}^n E((X^{(u)})^2) +\frac{2}{3}\Delta}\right). 
\end{aligned}
\end{equation}
For general independent random variables $|Y^{(u)}|\leq M$ $(u=1,2,\cdots,n)$, we need a normalization to apply Bernstein's inequality. If $\{Y^{(u)}\}$ is i.i.d., we can denote the expectation $E((Y^{(u)})^2)$ as $e$, then 
\begin{equation}\label{eq:devbernstein}
\begin{aligned}
&\sum_{u=1}^n Y^{(u)} \leq n E(Y^{(u)}) +\Delta(n, M, e, \epsilon) \\
&\sum_{u=1}^n Y^{(u)}  \geq n E(Y^{(u)}) -\Delta(n, M, e, \epsilon) \\
&\Delta(n, M, e ,\epsilon)  = -\frac{1}{3}M +\frac{1}{2}\sqrt{\left(\frac{2M \ln \epsilon}{3}\right)^2 -8n e \ln \epsilon}.
\end{aligned}
\end{equation} 
\end{theorem}

In our case we consider the i.i.d. random variable $\chi_P^{(u)}$, whose expectation is $E(\chi_P^{(u)})=\sum_{k=1}^{64} \lambda^*_k p_k$. The range is given by $M = \max(|\omega_{\mathrm{max}}|,|\omega_{\mathrm{min}}|)$. We can also calculate $E((\chi_P^{(u)})^2)$ with the values $\omega_w$. When the protocol parameters are given, the probability distribution of $\chi_P^{(u)}$ is fixed, then $M$ and $E((\chi_P^{(u)})^2)$ are also fixed. This enables us to define $\Delta_{\mathrm{Bern}}(N_{\mathrm{tot}},\epsilon) = \Delta(N_{\mathrm{tot}}, M, e, \epsilon)$ for given protocol parameters.
According to Eq.~\eqref{eq:devbernstein}, we have
\begin{equation}
\begin{aligned}
 \sum_{u=1}^{N_{\mathrm{tot}}}\chi_P^{(u)} & \leq N_{\mathrm{tot}} \sum_{k=1}^{64} \lambda^*_k p_k +\Delta_{\mathrm{Bern}}(N_{\mathrm{tot}},\epsilon) \\
\sum_{u=1}^{N_{\mathrm{tot}}}\chi_P^{(u)} & \geq N_{\mathrm{tot}} \sum_{k=1}^{64} \lambda^*_k p_k -\Delta_{\mathrm{Bern}}(N_{\mathrm{tot}},\epsilon),
\end{aligned}
\end{equation}
where each inequality holds with probability at least $1-\epsilon$.

\section{Primal SDP problem for the simplified phase-matching quantum key distribution}\label{app:SDP}

We define a new basis $\ket{\kappa_a, \kappa_b, \pi, \xi}_{A_0B_0C\Xi} \coloneqq \ket{\kappa_a}_{A_0} \ket{\kappa_b}_{B_0} \ket{\pi}_{A_1} \ket{\pi}_{B_1} \ket{\xi}_\Xi$. Then the phase error operator in Eq.~\eqref{eq:phaseoperator} can be expressed as
\begin{widetext}
\begin{equation}
\begin{aligned}
\hat{E}_{\mathrm{ph}}& = \frac{p_{\mathrm{aux},0}p_{\mathrm{basis},0}^2}{8}\sum_{\xi=0}^1 \sum_{\kappa_a=0}^1 \sum_{\kappa_b=0}^1 \ket{\kappa_a,\kappa_b,0,\xi}\bra{\kappa_a,\kappa_b,0,\xi}_{A_0B_0C\Xi}  \\
 &+ \frac{p_{\mathrm{aux},0}p_{\mathrm{basis},0}^2}{8}\sum_{\substack{\kappa_a,\kappa_b, \kappa_a^\prime,\kappa_b^\prime \in\{0,1\},\\ \kappa_a\neq \kappa_a^\prime, \kappa_b \neq \kappa_b^\prime }} \sum_{\xi=0}^1 (-1)^{\kappa_a+\kappa_b+\xi}\ket{\kappa_a,\kappa_b,0,\xi}\bra{\kappa_a^\prime,\kappa_b^\prime,0,\xi}_{A_0B_0C\Xi}.
\end{aligned}
\end{equation} 
\end{widetext}

The primal SDP is given by
\begin{widetext}
\begin{equation}\label{eq:PMQKDprimalre}
\begin{aligned}
&\max_{\hat{G}} \mathrm{tr}(\hat{E}_{\mathrm{ph}}\hat{G}) \\
\mathrm{s.t.}  & \quad \hat{G} \succeq 0 \\
&  \sum_{\xi=0}^1 \mathrm{tr}\left[\frac{1}{4}\left(\ket{0,1,0,\xi}\bra{0,1,0,\xi}_{A_0B_0C\Xi}+\ket{1,0,0,\xi}\bra{1,0,0,\xi}_{A_0B_0C\Xi}\right)\hat{G}\right] = \mathrm{tr}(\hat{E}_{\mathrm{bit}}^X\hat{G} )=p_{\mathrm{pass}}^{X,\mathrm{nom}} e_{\mathrm{bit}}^{X,\mathrm{nom}} \\
& \sum_{\xi=0}^1 \mathrm{tr}\left[\frac{1}{4}\left(\ket{0,1,1,\xi}\bra{0,1,1,\xi}_{A_0B_0C\Xi}+\ket{1,0,1,\xi}\bra{1,0,1,\xi}_{A_0B_0C\Xi}\right)\hat{G}\right]=  \mathrm{tr}(\hat{E}_{\mathrm{bit}}^Y\hat{G} )=p_{\mathrm{pass}}^{Y,\mathrm{nom}} e_{\mathrm{bit}}^{Y,\mathrm{nom}} \\
& \sum_{\xi=0}^1 \sum_{\kappa_a=0}^1 \sum_{\kappa_b=0}^1 \mathrm{tr}\left(\frac{1}{4}\ket{\kappa_a,\kappa_b,0,\xi}\bra{\kappa_a,\kappa_b,0,\xi}_{A_0B_0C\Xi}\hat{G}\right)=\mathrm{tr}(\hat{P}_{\mathrm{pass}}^X\hat{G}) =p_{\mathrm{pass}}^{X,\mathrm{nom}} \\
&  \sum_{\xi=0}^1 \sum_{\kappa_a=0}^1 \sum_{\kappa_b=0}^1 \mathrm{tr}\left(\frac{1}{4}\ket{\kappa_a,\kappa_b,1,\xi}\bra{\kappa_a,\kappa_b,1,\xi}_{A_0B_0C\Xi}\hat{G}\right) =\mathrm{tr}(\hat{P}_{\mathrm{pass}}^Y\hat{G}) =p_{\mathrm{pass}}^{Y,\mathrm{nom}} \\
& \mathrm{tr}(\hat{P}_{k} \otimes \hat{I}_\Xi \hat{G}) = p_{k} \\
& \hat{P}_k \otimes \hat{I}_\Xi= \left\{ 
\begin{array}{rcl}
\frac{1}{2}\sum_{\xi=0}^2\left(\ket{\kappa_a, \kappa_b,\pi,\xi}\bra{\kappa_a^\prime, \kappa_b^\prime,\pi^\prime,\xi}_{A_0B_0C\Xi}+ \ket{\kappa_a^\prime, \kappa_b^\prime,\pi^\prime,\xi}\bra{\kappa_a, \kappa_b,\pi,\xi}_{A_0B_0C\Xi}\right), \\
 4(\kappa_a+2\pi)+(\kappa_b+2\pi) \geq 4(\kappa_a^\prime+2\pi^\prime)+(\kappa_b^\prime+2\pi^\prime)\\
\frac{1}{2i}\sum_{\xi=0}^2\left(\ket{\kappa_a, \kappa_b,\pi,\xi}\bra{\kappa_a^\prime, \kappa_b^\prime,\pi^\prime,\xi}_{A_0B_0C\Xi}- \ket{\kappa_a^\prime, \kappa_b^\prime,\pi^\prime,\xi}\bra{\kappa_a, \kappa_b,\pi,\xi}_{A_0B_0C\Xi} \right),\\
4(\kappa_a+2\pi)+(\kappa_b+2\pi) < 4(\kappa_a^\prime+2\pi^\prime)+(\kappa_b^\prime+2\pi^\prime)
\end{array} \right. \\
& p_k = \left\{ 
\begin{array}{rcl} 
\mathrm{Re}\left(\braket{(-1)^{\kappa_{a}} (i)^{\pi} \sqrt{\mu_{\pi}}}{(-1)^{\kappa_{a}^\prime} (i)^{\pi^\prime} \sqrt{\mu_{\pi^\prime}} }
 \braket{(-1)^{\kappa_{b}} (i)^{\pi} \sqrt{\mu_{\pi}}}{(-1)^{\kappa_{b}^\prime} (i)^{\pi^\prime} \sqrt{\mu_{\pi^\prime}} }\right), \\
  4(\kappa_a+2\pi)+(\kappa_b+2\pi) \geq 4(\kappa_a^\prime+2\pi^\prime)+(\kappa_b^\prime+2\pi^\prime)\\
\mathrm{Im}\left(\braket{(-1)^{\kappa_{a}} (i)^{\pi} \sqrt{\mu_{\pi}}}{(-1)^{\kappa_{a}^\prime} (i)^{\pi^\prime} \sqrt{\mu_{\pi^\prime}} }
 \braket{(-1)^{\kappa_{b}} (i)^{\pi} \sqrt{\mu_{\pi}}}{(-1)^{\kappa_{b}^\prime} (i)^{\pi^\prime} \sqrt{\mu_{\pi^\prime}} }\right), \\
  4(\kappa_a+2\pi)+(\kappa_b+2\pi) < 4(\kappa_a^\prime+2\pi^\prime)+(\kappa_b^\prime+2\pi^\prime),\\
\end{array} \right. \\
\end{aligned}
\end{equation}
\end{widetext}
where $k=1+\kappa_a+2\pi+2^2\kappa_b+2^3(\kappa_a^\prime+2\pi^\prime+2^2\kappa_b^\prime)$ and $e_{\mathrm{bit}}^{X,\mathrm{nom}}$, $e_{\mathrm{bit}}^{Y,\mathrm{nom}}$, $p_{\mathrm{pass}}^{X,\mathrm{nom}}$ and $p_{\mathrm{pass}}^{Y,\mathrm{nom}}$ are estimations given in Appendix~\ref{app:simulation}. The factor $1/4$ comes from the assumption that Alice and Bob choose their bits $\kappa_{a(b)}\in\{0,1\}$ with equal probability.

\section{Simulation of the finite-size key rate}\label{app:simulation}

In this section we simulate the key rate of the simplified PM-QKD protocol given in the last section.
We need to estimate $e_{\mathrm{bit}}^{X,\mathrm{nom}}$, $e_{\mathrm{bit}}^{Y,\mathrm{nom}}$, $p_{\mathrm{pass}}^{X,\mathrm{nom}}$ and $p_{\mathrm{pass}}^{Y,\mathrm{nom}}$.
Here we consider a simple lossy channel model. Suppose the dark count rate of the single photon detector is $p_d$ and the channel transmittance is $\eta$.
We make the following estimations,
\begin{equation}\label{eq:simulation}
\begin{aligned}
e^{X,\mathrm{nom}}_{\mathrm{bit}} & =e^{-2\eta \mu_1}p_d(1-p_d) \\
e^{Y,\mathrm{nom}}_{\mathrm{bit}} & =e^{-2\eta \mu_2}p_d(1-p_d) \\
p^{X,\mathrm{nom}}_{\mathrm{pass}} & =(1-p_d)\left[1-e^{-2\eta \mu_1}(1-p_d)\right]+e^{-2\eta \mu_1}p_d(1-p_d) \\
p^{Y,\mathrm{nom}}_{\mathrm{pass}} & =(1-p_d)\left[1-e^{-2\eta \mu_2}(1-p_d)\right]+e^{-2\eta \mu_2}p_d(1-p_d),\\
\\
\end{aligned}
\end{equation}
where the loss and double click events are regarded as the inconclusive results. Then $N_{\mathrm{sig}}^{\mathrm{nom}}$ and $\Theta_{Q,l}^{\mathrm{nom}}$ are given in terms of the protocol parameters,
\begin{equation}
\begin{aligned}
N_{\mathrm{sig}}^{\mathrm{nom}} & = N_{\mathrm{tot}}(1-p_{\mathrm{trash}})p_{\mathrm{basis,0}}^2 p_{\mathrm{aux},0} p^{X,\mathrm{nom}}_{\mathrm{pass}} \\
\Theta_{Q,1}^{\mathrm{nom}} &= N_{\mathrm{tot}}(1-p_{\mathrm{trash}})p_{\mathrm{basis,0}}^2 p_{\mathrm{aux},1} p^{X,\mathrm{nom}}_{\mathrm{pass}} e^{X,\mathrm{nom}}_{\mathrm{bit}}\\
 \Theta_{Q,2}^{\mathrm{nom}} &= N_{\mathrm{tot}}(1-p_{\mathrm{trash}})p_{\mathrm{basis,1}}^2  p^{Y,\mathrm{nom}}_{\mathrm{pass}} e^{Y,\mathrm{nom}}_{\mathrm{bit}}\\
\Theta_{Q,3}^{\mathrm{nom}} &= N_{\mathrm{tot}} (1-p_{\mathrm{trash}})p_{\mathrm{basis,0}}^2 p_{\mathrm{aux},1}p^{X,\mathrm{nom}}_{\mathrm{pass}} \\
\Theta_{Q,4}^{\mathrm{nom}} &= N_{\mathrm{tot}} (1-p_{\mathrm{trash}})p_{\mathrm{basis,1}}^2 p^{Y,\mathrm{nom}}_{\mathrm{pass}}.
\end{aligned}
\end{equation}

In order to calculate the finite-size key rate, we need both the actual experimental results and their estimations.
We assume all the actual experimental values equal their estimations, i.e., $N_{\mathrm{sig}}=N_{\mathrm{sig}}^{\mathrm{nom}}$ and $\Theta_{Q,l}=\Theta_{Q,l}^{\mathrm{nom}}$ $(l\in\{1,2,3,4\})$.
We set $\epsilon_j = 2^{-66}/14$ for $j\in\{0,1,2,3,4,6,7\}$. Then $\epsilon_5=2^{-66}/7$ and $\epsilon_{\mathrm{ph}}=2^{-66}$ according to the union bound. We also set $s=66$ and $s^\prime=32$. The security parameter is given by $\epsilon_{\mathrm{tot}}=\sqrt{2}\sqrt{\epsilon_{\mathrm{ph}}+2^{-s_{\mathrm{PA}}}}+2^{-s^\prime}=2^{-31}<10^{-10}$. The error correction efficiency is $1.1$.

\bibliography{bibfinite}
\bibliographystyle{apsrev4-1}
\end{document}